\documentclass[11pt]{article}

\pdfoutput=1
\usepackage{cite}
\usepackage{hyperref}
\usepackage{allrunes}
\usepackage{epiolmec}
\usepackage[dvipsnames]{xcolor}

\usepackage{lineno}

\usepackage{amsmath}
\usepackage{amssymb,amsfonts,amsthm}
\usepackage{epsfig,graphicx}
\usepackage{url} 
\usepackage{bm} 
\usepackage{slashed}
\usepackage{cite}

\usepackage{color}
\usepackage{pstricks}

\usepackage{fancyhdr}
\usepackage{textcomp}

\usepackage{xcolor} 
\usepackage{colortbl} 
 
\newcommand{\unit}{\leavevmode\hbox{\small1\kern-3.6pt\normalsize1}}

\def\mdm{m_{\chi}}

\def\locald{\rho_0}

\def\mnucl{m_{T}}

\def\redT{\mu_T}

\newcommand{\acro}{{\sc RAPIDD}}

\newcommand{\mwimp}{m_\chi}

\def\op#1{{\cal O}_{#1}}

\begin{document}

\title{RAPIDD}

\thispagestyle{empty}
\begin{flushright}

 {\small 
 IPPP/18/12; DCTP/18/24
 \\
 FERMILAB-PUB-18-047-A-AE-CD
 }

  \vspace*{2.mm}{February 8, 2018}
\end{flushright}

\begin{center}
  {\bf {\LARGE Surrogate Models for Direct Dark Matter Detection}}

\renewcommand*{\thefootnote}{\fnsymbol{footnote}}
\setcounter{footnote}{3}

  \vspace{0.5cm}
  {\large
    D.~G.~Cerde\~no $^{a}$,
    A.~Cheek $^{a}$,
    E.~Reid $^{a}$, and
    H.~Schulz $^{b}$
  }
  \\[0.2cm]

  {\footnotesize{
$^a$ Institute for Particle Physics Phenomenology, Department of Physics\\
Durham University, Durham DH1 3LE, United Kingdom \\
$^b$ Department of Physics, University of Cincinnati, Cincinnati, OH 45219, USA

Email: davidg.cerdeno@gmail.com, andrew.cheek@durham.ac.uk, elliott.m.reid@durham.ac.uk, iamholger@googlemail.com
        }
    }

\vspace*{0.7cm}

  \begin{abstract}
In this work we introduce RAPIDD, a surrogate model that speeds up the computation of the expected spectrum of dark matter particles in direct detection experiments. RAPIDD replaces the exact calculation of the dark matter differential rate (which in general involves up to three nested integrals) with a much faster parametrization in terms of ordinary polynomials of the dark matter mass and couplings, obtained in an initial training phase. In this article, we validate our surrogate model on the multi-dimensional parameter space resulting from the effective field theory description of dark matter interactions with nuclei, including also astrophysical uncertainties in the description of the dark matter halo. As a concrete example, we use this tool to study the complementarity of different targets to discriminate simplified dark matter models. We demonstrate that RAPIDD is fast and accurate, and particularly well-suited to explore a multi-dimensional parameter space, such as the one in effective field theory approach, and scans with a large number of evaluations.
\end{abstract}
\end{center}

\newpage


\section{Introduction}
\label{sec:introduction}
\setcounter{equation}{0}


Astrophysical and cosmological observations strongly indicate that approximately 85\% of the matter density of the Universe consists of a new type matter that does not emit or absorb light. The detection and identification of this dark matter (DM) constitutes one of the main challenges in modern particle physics, as it can only be explained with new physics beyond the Standard Model. Among the different particle models for DM, a generic weakly-interacting massive particle is considered a natural candidate since it can be thermally produced in the early Universe in the right amount to account for the observed DM abundance today. DM particles with electroweak scale interactions can be searched for directly, through their scattering off nuclei in underground detectors. A large number of experiments have been looking for the resulting keV-scale nuclear recoils using a variety of techniques during the past decades. No confirmed DM signature has been found, which has lead to stringent upper constraints on the DM-nucleus scattering cross section. In the coming years, a new generation of detectors will continue probing the DM paradigm with improved sensitivities and larger targets, raising the hope of a future detection.

The expected event rate from the elastic scattering of a DM particle, ${\chi}$, with mass $\mwimp$ off a target nucleus with mass $\mnucl$ in a given energy bin, $k$, is given by
\begin{equation}
N_k =\frac{\locald \epsilon}{\mnucl\,\mwimp}\int_{E_k}^{E_{k+1}}dE_R\,\varepsilon(E_R) \int_{E^\prime_R}dE^\prime_R\,Gauss(E^\prime_R, E_R) \int_{v_{min}} d \vec v\, v f(\vec v)\, \frac{d\sigma_{{\chi}T}}{dE^\prime_R}  \,,\label{eq:drate}
\end{equation}
where $\locald$ is the local DM density, $f(\vec v)$ is the DM velocity distribution in the detector frame normalized to unity, and $\epsilon$ is the total exposure (given by the product of the  detector mass 
and the run-time). 
The integration over the DM velocity is performed from the minimum DM speed needed to induce a nuclear recoil of energy $E_{R}$, $v_{min}=\sqrt{\mnucl E_R/(2\redT^2)}$.
The total event rate is  then calculated by integrating the differential event rate over the nuclear recoil energy, $E_R$, within a given energy bin, dependent on the specific experiment (bins are defined from a minimum threshold energy $E_{T}$). In doing this, the experimental energy resolution, $Gauss(E_R^\prime,E_R)$ (generally incorporated as a Gaussian smearing), and energy-dependent efficiency, $\varepsilon(E_R)$, have to be taken into account.

The particle physics nature of the DM is encoded in the DM-nucleus differential scattering cross section, $d\sigma_{{\chi}T}/dE_R$. Traditionally, the scattering cross-section is split into two components, a spin-dependent (SD) and a spin-independent (SI) one, which originate from different terms in the microscopic Lagrangian describing DM interactions with quarks. However, the general Lagrangian describing DM interactions with nuclei in the non-relativistic limit can be much more diverse \cite{Fan:2010gt}, featuring up to 18 different operators, some of which display a non-trivial dependence with the DM velocity and the momentum exchange \cite{Fan:2010gt,Fitzpatrick:2012ix,Dent:2015zpa}. The resulting effective field theory (EFT)  is then described in terms of a Lagrangian that contains four-field operators of elastic scattering between a dark matter particle and a target nucleon,
\begin{equation}
{\cal L}_{\text{int}}=\sum_{\tau} \sum_i c_i^{\tau}\mathcal{O}_i \overline{\chi} \chi \overline{\tau} \tau.
\label{eq:eft}
\end{equation}
In this expression, $\tau$ can either represent proton and neutron interactions or isoscalar and isovector interactions. In our work we will use the isospin basis, thus $\tau=0,1$. The operator variables of the effective Lagrangian must be invariant under Galilean transformations. This means that the momentum- and velocity-dependent terms must appear as the momentum transfer and the relative incoming velocities, which limits the number of effective operators \cite{Fitzpatrick:2012ix}. The total DM-nucleus cross section is calculated by adding these contributions coherently, using nuclear wave functions, which results in the following expression,
\begin{equation}
\frac{d\sigma_{{\chi}T}}{dE_R}=\frac{m_T}{2\pi m_v^4} 
\frac{1}{v}\sum_{ij}\sum_{\tau, \tau'=0,1} c^{\tau}_ic^{\tau'}_j\mathcal{F}^{\,\tau,\tau'}_{i,j}(v^2,q^2)
\ .
\label{eq:dsigma}
\end{equation}
Here, $\mathcal{F}_{i,j}^{\,\tau,\tau^\prime}$ are the nuclear form factors (see e.g., Refs.\,\cite{Fitzpatrick:2012ix,Anand:2013yka} for their expressions in the isospin and nuclear basis, respectively). This expression explicitly shows the occurrence of interference terms between the two isospin-components within each operator, $\mathcal{F}_{i,i}^{\,0,1}$, (relevant when the coupling of the DM to protons and neutrons differs), as well as interference terms between the following pairs of EFT operators, ($\op{1},\op{3}$), ($\op{4},\op{5}$), ($\op{4},\op{6}$), and ($\op{8},\op{9}$). The couplings $c_i$ are chosen to be dimensionless, having been normalised by the Higgs vacuum expectation value, $m_v=264$~GeV, following the prescription of Ref.\,\cite{Anand:2013yka,Catena:2014epa}.

In the case of a positive signature, the spectral shape of the nuclear recoil spectrum can be used to reconstruct the dark matter properties \cite{Green:2007rb,Green:2008rd}. In fact, the shape of the spectrum contains information that allows to distinguish non-standard momentum dependent contributions \cite{McDermott:2011hx}. The reconstruction of DM parameters is subject to statistical limitations \cite{Strege:2012kv} and is also very sensitive to uncertainties in the astrophysical parameters describing the Milky Way halo \cite{Green:2011bv}, as well as in the nuclear form factors \cite{Cerdeno:2012ix}. Finally, it has been shown that the use of different experimental targets \cite{Bertone:2007xj,Pato:2010zk,Cerdeno:2013gqa} is crucial in order to determine the DM parameters after a positive detection. The reconstruction of DM parameters is extremely challenging in the multi-dimensional EFT parameter space. Combining the results from multiple targets and techniques strongly constrains theoretical models in the absence of a detection and allows for determination of the underlying physics of the interaction once a signal is seen \cite{Peter:2013aha,Catena:2014uqa,Kahlhoefer:2017ddj}.  It has thus been argued that next generation experiments constitute an excellent tool to  probe the general EFT parameter space \cite{Gluscevic:2014vga,Catena:2014epa} and identify the right theory \cite{Catena:2014hla,Gresham:2014vja,Gluscevic:2015sqa,Queiroz:2016sxf,Kavanagh:2017hcl}. Adding information from annual modulation \cite{Witte:2016ydc} is particularly useful to identify certain class of unconventional operators. In Ref.\,\cite{Rogers:2016jrx} a strategy to explore the vast EFT parameter space using direct detection data was tested, based on the use of Bayesian inference methods. Finally, the inclusion of data from indirect searches and colliders (LHC) provides very valuable complementary information with which the DM properties can be better determined (see e.g., Refs.\cite{Bernal:2008zk,Bertone:2010rv,Bertone:2011pq,Frandsen:2012rk, Mambrini:2012ue,Arbey:2013iza,Liem:2016xpm,Baum:2017kfa,Bertone:2017adx}).

From a computational perspective, the reconstruction of DM parameters involves evaluating equation (\ref{eq:drate}) in a multidimensional parameter space, which can be very costly since in general it contains three nested integrals. In order to speed up this process, we have developed \acro, a surrogate model that allows a fast and accurate determination of the expected DM spectrum in direct detection experiments. In particular, we have used the {\sc Professor} tool \cite{Buckley:2009bj} to parameterise the experimental response of  direct dark matter experiments in terms of simple polynomial functions. The polynomial fits are obtained for each individual experiment via a training process, which employs the exact calculation and the specific details of each experiment. After this (expensive) offline phase, the resulting surrogate model is considerably faster than the exact calculation, especially when the dimensionality of the parameter space is large. Thus, it is ideal to explore the general EFT parameter space, to investigate the complementarity of different targets, or to use in scans that require a large number of evaluations. In this article we validate \acro, and we use it to test the identification of simplified DM models using direct detection data from upcoming experiments.

This article is organised as follows. In Section~\ref{sec:numerical}, we explain how the surrogate model \acro\ is built. We comment on possible limitations and explain how these are dealt with in our analysis. In Section~\ref{sec:examples} we test \acro\ in some simple scenarios, based on one and two effective operators to describe the DM-nucleus scattering cross section, and also including astrophysical uncertainties. To illustrate our method, in Section~\ref{sec:simplified}, we apply it to study the reconstruction of parameters with a simplified model approach, which involves up to four different operators, employing three different experimental targets. Finally, our conclusions are presented in Section~\ref{sec:conclusions}.

\section{Parametrization of the DM detection rate}
\label{sec:numerical}
\setcounter{equation}{0}

In this section, we explain the construction of a surrogate model to compute the expected number of DM events in direct detection experiments. Our goal is to speed up the computation without losing precision, in order to adapt it to explore multi-dimensional parameter spaces and large scans. To this aim, we have developed \acro\ (Reconstruction Algorithm of Parameters In Direct Detection), a Python code based on the {\sc Professor} tool (used extensively in particle collider analyses).

The idea of replacing the expensive part of a calculation with an approximate model is certainly not new. In fact a very similar approach taken in this work has been successfully applied in the field of collider physics to optimise parameters of Monte-Carlo programs using numerical $\chi^2$ minimisation \cite{Abreu:1996na} or constrain to effective field theory operators in BSM physics scenarios \cite{Englert:2017aqb,Buckley:2015lku,Englert:2015hrx}.

The objective function of our optimization problem is a binned likelihood, $\mathcal{L}(\mathbf{\Theta})$, constructed from a signal prediction and data (we will use mock data to simulate hypothetical future results):
\begin{equation} \mathcal{L}(\mathbf{\Theta}) = \prod_a
    \mathcal{L}^a(\mathbf{\Theta}) = \prod_a\prod_k{
        \frac{N_k^a(\mathbf{\Theta})^{\lambda_k^a}
        e^{N_k^a(\mathbf{\Theta})}}{\lambda_k^a!}} \,, \label{eqn:likelihood}
\end{equation} 
where the number of recoil events $N_k^a$ in the $k$-th bin for the experiment
$a$, is compared with the prediction of the benchmark model in the same bin,
$\lambda_k^a$, for that given target. We assume that each experimental dataset
follows an independent Poissonian distribution, so the full likelihood function is
equivalent to the product of the likelihoods for each experiment
$\mathcal{L}^a$.

Traditionally, such analyses are conducted by interfacing a likelihood
evaluator with the signal generating code directly. Modern statistical tools
are very efficient and especially the introduction of nested sampling prevents
wasting CPU cycles on points of the parameter space where the likelihood is
low. However, this approach relies on exactly evaluating the signal prediction
at every iteration which in turn means that the run-time of the likelihood
evaluation is dominated by the run-time of the signal prediction of
eq.\,(\ref{eq:drate}) (which in general involves three nested integrals). This
effectively limits its applicability to low-dimensional parameter spaces or
more general to cases where the statistical analysis does not become
prohibitively expensive.

\begin{figure}[t!]
\begin{center}
\includegraphics[scale=0.6, bb=0 0 432 288]{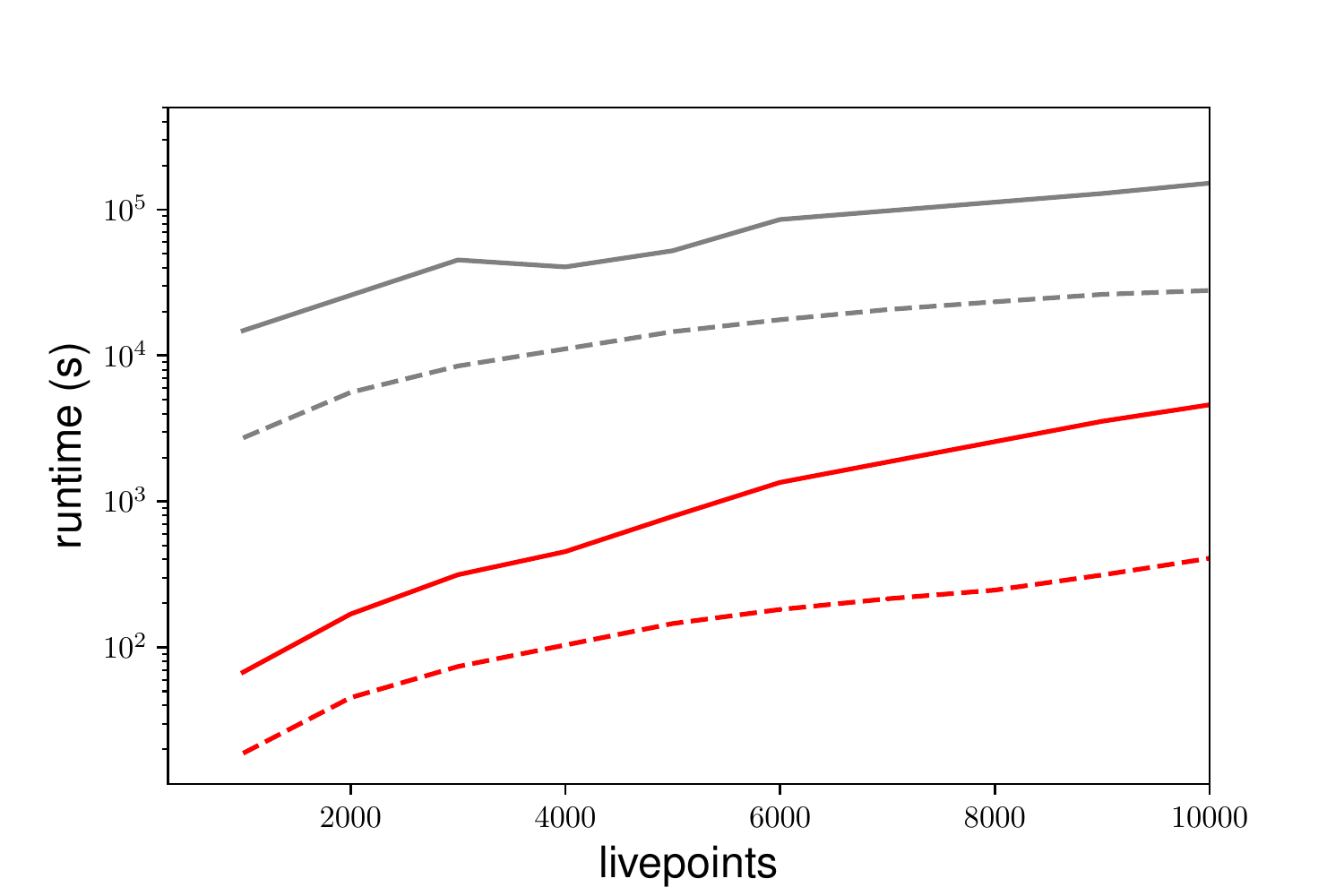}
\end{center}
\caption{Execution time of the surrogate model \acro\ (red lines) as a function of the number livepoints used in MultiNest runs. For reference, the gray lines represent the runtime of the full physics code. The solid (dashed) lines correspond to the case without (with) astrophysical uncertainties.}
\label{fig:speed}
\end{figure}

In this work we replace each exact $N_k^a$ with an ordinary polynomial that has
been trained on a sample of $N_k^a$  at various points of the model parameter
space, $\mathbf{\Theta}$, using {\sc Professor}. In that sense, our surrogate
model is simply a collection of polynomials and the computational gain is
due to the polynomials being much cheaper to evaluate than the true $N_k^a$
in eq.\,(\ref{eqn:likelihood}). 
In Figure~\ref{fig:speed} we compare the execution time of our surrogate model with that of the full physics code. To evaluate the Bayesian evidence in both cases, we will use MultiNest \cite{Feroz:2008xx,Feroz:2007kg}, which is particularly well suited to explore high dimensional parameter spaces with multi-modal posterior distributions. The execution time is represented as a function of the number of livepoints used in MultiNest. We have observed a consistent improvement of approximately two orders of magnitude in the speed of the computation in the simplest runs with a small number of parameters.

It goes without saying that the method allows for other parametrization
functions. We choose polynomials, however, as they are numerically robust, easy
to understand and relatively cheap to train. Their usage is further motivated
by the fact that the number of DM events in a given energy bin, given by equation
(\ref{eq:drate}), is in general a smooth function of the DM parameters (mass
and couplings) in a given energy range.  There are exceptions to this mild
behaviour that will require a more careful treatment, namely accidental
cancelations due to interference terms between different operators, and
threshold effects for low DM masses. We briefly summarise {\sc Professor} here
before addressing these points.

The objective of {\sc Professor} is to translate the exact signal prediction in equation~(\ref{eq:drate}) for each bin, $N_k^a(\mathbf{\Theta})$, into an ordinary polynomial, ${\cal P}_k^a(\mathbf{\Theta})$.
To do so we first choose a polynomial order $\mathcal{O}$ appropriate for the physics problem at hand. With $\mathcal{O}$ and $\mathbf{\Theta}$ given, the structure of the polynomial is fixed. What remains to be done is to determine the $N_\mathrm{coeffs}$ coefficients, $d^{a}_{k,l}$, that allow to approximate the true behaviour of $N_k^a(\mathbf{\Theta})$ such that
\begin{equation}
N_k^a(\mathbf{\Theta}) \approx {\cal P}_k^a(\mathbf{\Theta}) = \sum_{l=1}^{N_\mathrm{coeffs}} d^{a}_{k,l} \, \tilde{\Theta}_l  \equiv \mathbf{d^{a}_k} \cdot \mathbf{\tilde{\Theta}},
\label{eq:param}
\end{equation}
where $\tilde{\Theta}_l$ are suitable combinations of the actual parameters\footnote{
For example, for a quadratic polynomial in a two dimensional parameter space $\mathbf{\Theta}=(\mdm,\,c_1)=(x, y)$,  
these would take on the form
$\mathbf{d^{a}_k} = (\alpha, \beta_x,\beta_y, \gamma_{xx}, \gamma_{xy}, \gamma_{yy}),
$
and
$\mathbf{\tilde{\Theta}} = (1, x, y, x^2, xy, y^2).
$}, $\mathbf{\Theta}$.
The algorithm to determine $\mathbf{d^a_k}$ requires the knowledge of the exact $N_k^a(\mathbf{\Theta})$ at some (randomly sampled) points  of the parameter space in question. Having sampled and evaluated $N_k^a$ for at least $N_\mathrm{coeffs}$
points (a task that can trivially be parallelised) we can construct a matrix equation
\begin{equation}
\vec{N_k^a} =M_{\mathbf{\tilde{\Theta}}}\cdot \mathbf{d^{a}_k},
\label{eq:inversion}
\end{equation}
where $M_{\mathbf{\tilde{\Theta}}}$ is a quantity similar to a Vandermorde matrix where each row
contains the values of $\mathbf{\tilde{\Theta}}$ for each sampled point, and
$\vec{N_k^a}$ is a vector of the resulting number of events.  This allows us to
solve for $\mathbf{d^{a}_k}$ using the (pseudo-) inverse of
$M_{\mathbf{\tilde{\Theta}}}$, which in the {\sc Professor} program is
evaluated by means of a singular value decomposition.

The minimal number of points (i.e. fully determined matrix) is given by the number of coefficients of an r-th order polynomial in D dimensions. The exact number is given in \cite{Buckley:2009bj}. We found it beneficial to oversample by approximately a factor of 2 in order to have greater statistics when validating our parameterization.

Although extremely robust and justified whenever Taylor's theorem applies, the validity of the polynomial approximation is not guaranteed and must be checked before attempting any likelihood evaluation. For the most part, standard techniques such as checking the polynomial prediction against its own exact inputs provided by the {\sc Professor} toolkit were used. In this work specifically we were confronted with the following limitations:

 \begin{enumerate} 
\item[(i)]  Low-mass DM: 
The number of expected DM events for a given energy bin is in general a smooth function of the DM mass (and therefore susceptible to be fit by a polynomial). The only subtlety to take into account is that, for a given DM mass, there is a maximum recoil energy, given by
\begin{equation}
E_R^{max}=2\frac{\mu_{\chi T}^2}{m_T} v_{esc}^2\ ,
\label{eq:emax}
\end{equation}
where $v_{esc}$ is the escape velocity in the DM halo. 
If the incident particle is light enough, experiments will be able to probe the end point of the spectrum, which means that $N_k^a$ is zero above a given energy bin.
In our parametrisation, this discontinuity is difficult to fit precisely with a polynomial function. We have circumvented this difficulty by multiplying by a Heaviside step function which automatically incorporates condition (\ref{eq:emax}).

\item[(ii)] Accidental cancellations: as already mentioned in the introduction, there are interference terms between the different isospin contributions for each operator, as well as between some of the EFT operators. These subtleties are difficult to capture with the single polynomial approximation proposed in eq.\,(\ref{eq:param}). Instead, we have found that it is much more convenient to use various polynomials (one for each effective operator, including also the interference term), as follows
 
 \begin{equation}
N_k^a(\mathbf{\Theta}) \approx \sum_{ij}\sum_{\tau,\tau^\prime=0,1} {\cal P}_k^{a,i,j,\tau,\tau^\prime}(\mathbf{\Theta})
\, .
\label{eq:param2}
\end{equation}
Building the parametrisation in this way makes the training stage quicker, because the required number of sample points is reduced. Solving equation (\ref{eq:inversion}) for the coefficients in a lower dimension is also quicker than in a higher dimension, which compensates for building multiple polynomials for each dimension.

\item[(iii)] Precision loss: For consistency, we have monitored the precision of the surrogate model by comparing the DM spectrum obtained for the best fit point with the surrogate model and with the physics code. We have found that in general the agreement was excellent, well below 1\% for the examples shown in this paper. We have found that precision can be lost in some cases of high dimensionality, but that this behaviour can be corrected if a higher order in the polynomial fit is used. Likewise, the surrogate model can be less precise towards the edges of the parameter space used in the training phase. This is easily avoided by training the surrogate model in a wider window that the one where it is intended to be used. 
\end{enumerate}

\section{Examples}
\label{sec:examples}
\setcounter{equation}{0}

In this section we consider various simple examples that allow us to validate our surrogate model. We have selected various DM benchmark points that are within the reach of future G2 experiments and we have attempted to reconstruct the DM parameters (mass and couplings) using \acro, and comparing it with the full calculation.

\begin{table}[t!]
	\begin{center}
	\begin{tabular}{ |l | l| l| l|}
 \hline
	 Target&Exposure&Energy window & Bin No\\
	\hline
	 Xe&5.6$\times 10^6$ kg days& 3-30 keV& 27\\
	 Ge&91250 kg days&0.35-50 keV& 49\\
    Ar & 7.3$\times 10^6$ kg days & 5.0-30 keV& 24\\
	\hline
	\end{tabular}
	\end{center}
\caption{Specifications of the direct detection experiments considered in this work.}
\label{tab:experiments}
\end{table}

In Table\,\ref{tab:experiments}, we summarise the experimental configurations that we have considered in this work. These are motivated by future direct detection experiments. The exposure and energy ranges are chosen so as to mimic the planned G2 experiments SuperCDMS \cite{Agnese:2016cpb} (for Ge and Si), LZ, XENON1T, PandaX \cite{Akerib:2015cja,Aprile:2017aty,Cui:2017nnn} (for Xe) and DarkSide \cite{Aalbers:2016jon} (for Ar). Notice, however, that at this point we are not interested in replicating the whole experimental setup, and for simplicity we  also assume a constant efficiency, $\varepsilon(E_R)=1$, and perfect energy resolution, $Gauss(E_R^\prime,E_R)=\delta(E_R^\prime-E_R)$, in Eq.\,(\ref{eq:drate}). These quantities vary from experiment to experiment, and can be straightforwardly incorporated in our method, only having a cost in the initial training time.

The parameter reconstruction is carried out using MultiNest 2.9 \cite{Feroz:2008xx,Feroz:2007kg}, which is interfaced with \acro. 
In order to test the results with the full computation, we also interface MultiNest to our own numerical code that computes the number of recoil events using Eq.(\ref{eq:drate}). In both cases, we use the same definition for the likelihood, based on a binned analysis of the resulting data.
Scans are performed with 15000 live points and a tolerance of 0.0001 to reach a good sampling of the profile likelihood (defined below) as found in Ref.\,\cite{Feroz:2011bj}.

The experimental data consists of the predicted sets of binned DM rates for each target, $\mathbf{D}=(\{\lambda_k^a\})$. The parameter space is therefore $\mathbf{\Theta}=(\mdm,\,c_i^\tau)$. Logarithmic priors are assumed for the EFT couplings and for the DM mass. 
Regarding the properties of the DM halo, in the first examples we will consider the Standard Halo Model (SHM). The SHM is characterised by an isotropic Maxwell-Boltzmann velocity distribution function \cite{Lewin:1995rx} in Eq.\,\ref{eq:drate}. We have used the following values for the local dark matter density, $\locald=0.4$~GeV~cm$^{-3}$, central velocity, $v_0=220$~km~s$^{-1}$, and escape velocity, $v_{esc}=544$~km~s$^{-1}$.
In this first example, we have not incorporated uncertainties in these quantities, but we will address this in  Section~\ref{sec:haloes}, together with the generalisation to other DM haloes.

\subsection{One operator: spin-independent scattering}
In order to tune our method, we have started with a canonical scenario, where the DM-nucleus scattering cross section is described by a single operator. We have chosen $\op{1}$, which corresponds to the standard spin-independent scattering, this can come from a scalar, fermion dark matter particle as well as more exotic natures\footnote{In the next section we will explore specific realizations in terms of simplified DM models, where one can see that $\op{1}$ is ubiquitous. For this first test, there is no need to specify the model.}. We have also set the couplings of the DM to protons and neutrons equal ($c_i^p=c_i^n=c_i^0$), which is known as being purely isoscalar. We are therefore left with a two-dimensional parameter space $(\mdm,\,c_1^0)$.

We have chosen two benchmark points, a low mass case with $\mdm=30$~GeV and a higher mass case, where $\mdm=100$~GeV. The coupling to $\op{1}$ is $c_1^0=5\times10^{-5}$ in both cases. For reference, the relation with the (most commonly used) zero-velocity spin-independent DM-nucleon scattering cross-section reads
\begin{equation}
\sigma_{\chi N}= \frac{\mu_{\chi N}^2}{\pi\, m_{v}^4}\left(c^0_1\right)^2\ ,
\end{equation} 
where $\mu_{\chi N}$ is the DM-nucleon reduced mass and $m_{v}$ is the Higgs expectation which enters the calculation by normalization convention explained in section \ref{sec:introduction}. The parameterization was trained using $1000$ random points in the $(\mdm,\,c_1^0)$ plane.

\begin{figure}
\begin{center}
\includegraphics[scale=0.55, bb=0 0 460.8 345.6]{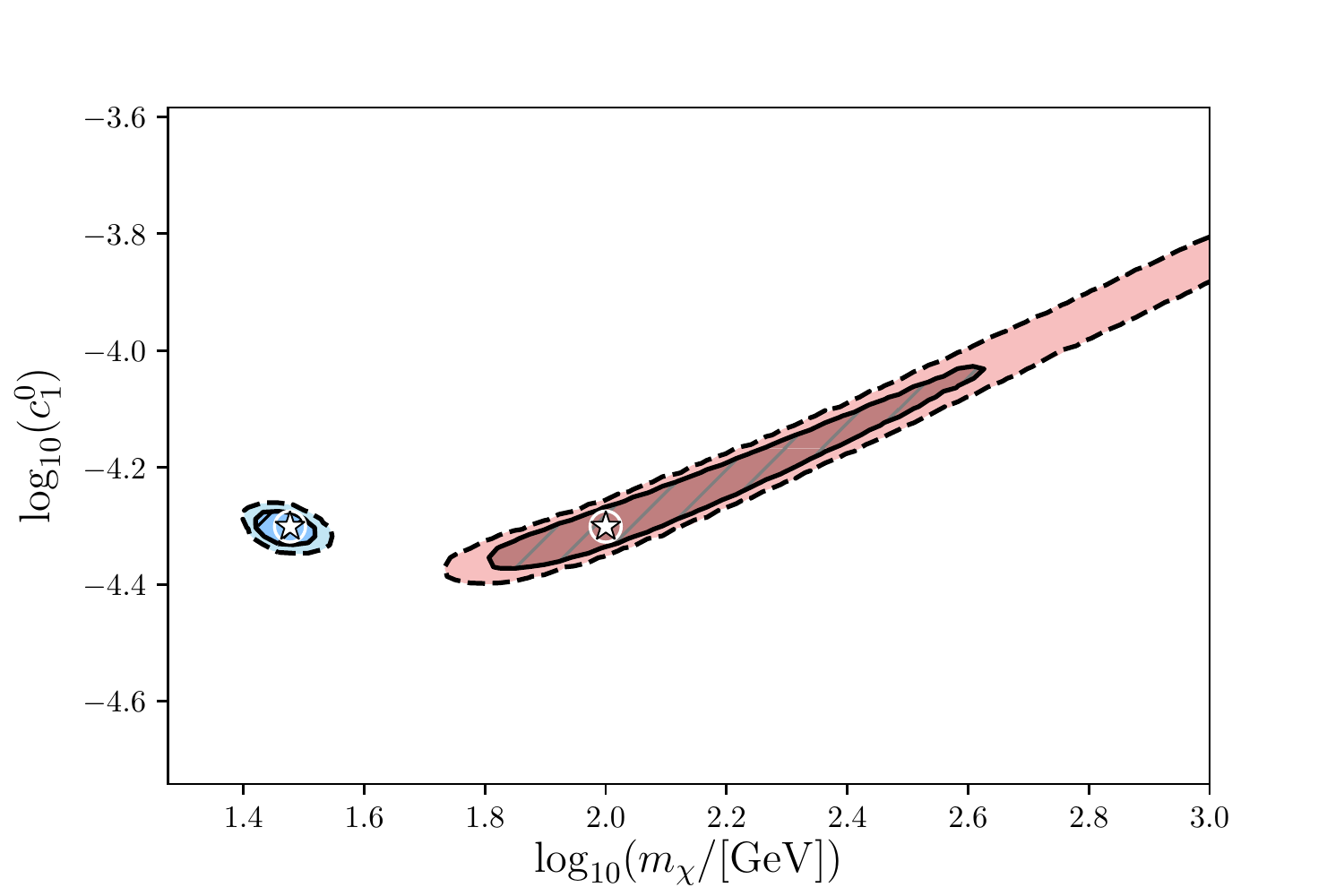}
\end{center}
\caption{Reconstruction of DM parameters in the $(\mdm,\,c_1^0)$ plane for two benchmark points. The best-fit point and $1\,\sigma$ (68\% C.L.) and $2\,\sigma$ (95\% C.L.) regions are shown for the results obtained with \acro\ (white star and shaded areas) and the full physics code (white ring and solid and dashed black lines).} 
%
\label{fig:comparison_o1}
\end{figure}

In Figure~\ref{fig:comparison_o1} we show the reconstruction of DM parameters for both benchmark points, the light mass, which is shaded blue, and the high mass which is shaded red. The black lines indicate the $2\sigma\,$ and $1\sigma$ regions calculated by the physics code. We have assumed observation in a future Xenon experiment (with details as in Table~\ref{tab:experiments}). The best fit points are, respectively, $(\mdm,\,c_1^0)=(30.0\, \textrm{GeV},\,5.00\times 10^{-5})$ and  $(99.7\,\textrm{GeV},\,4.99\times 10^{-5})$, and the $1\,\sigma$ (68\% C.L.) and $2\,\sigma$ (95\% C.L.) regions span the same areas. 
Without having lost accuracy, the great advantage of the parametrisation method is its speed.
While the full computation took approximately $40$ minutes for each example, the results using the surrogate model took just $10$ seconds (after an initial training phase of approximately 2 minutes). 
To calculate the contours in these plots we have used a function provided by Superplot \cite{Fowlie:2016hew}.

As mentioned in Section~\ref{sec:numerical}, low DM masses are a potential challenge for our surrogate model. With this test we have shown that \acro\ is reliable in this mass regime.

\subsection{Operator interference and isospin-violating couplings}
\label{sec:isospin}

As explained in the Introduction, each operator's response is summed over isoscalar and isovector interactions (or equivalently, proton and neutron interaction). Likewise, there are interference terms among some of the EFT operators. Due to the resulting interference terms, accidental cancellations can occur between these responses. For example, the interaction rate of isospin-violating dark matter
\cite{Kurylov:2003ra,Giuliani:2005my,Chang:2010yk,Feng:2011vu} 
is extremely sensitive to the nuclear target. In fact, for specific choices of DM couplings to protons and neutrons, one can greatly suppress the expected rate in certain targets, a strategy that was once used to try to reconcile positive DM hints (such as DAMA and CoGeNT) with the negative results from other experiments (mainly XENON). 
For a recent review on isospin-violating DM models, see Ref.\,\cite{Yaguna:2016bga}.

This finely tuned cancellation is a challenge for our parametrisation technique. In particular, we have checked that a polynomial approximation of the total response, $dR/dE_R$, is unable to properly capture this subtle behaviour. As already mentioned in Section \ref{sec:numerical}, this problem can be addressed by using independent parametrisations for each isospin contribution and for each interference term, as in equation (\ref{eq:param2}).
In this particular example, we will use different polynomials for each of the three contributions, ${\cal P}^{00}_k$, ${\cal P}^{10}_k$, and ${\cal P}^{11}_k$. Each of these vary smoothly with the input parameters $(\mdm,\,c_1^0,\,c_1^1)$ and this ensures a much more reliable reconstruction, including cancellations.

\begin{figure}
\begin{center} 
\includegraphics[scale=0.6, bb=0 0 460.8 345.6]{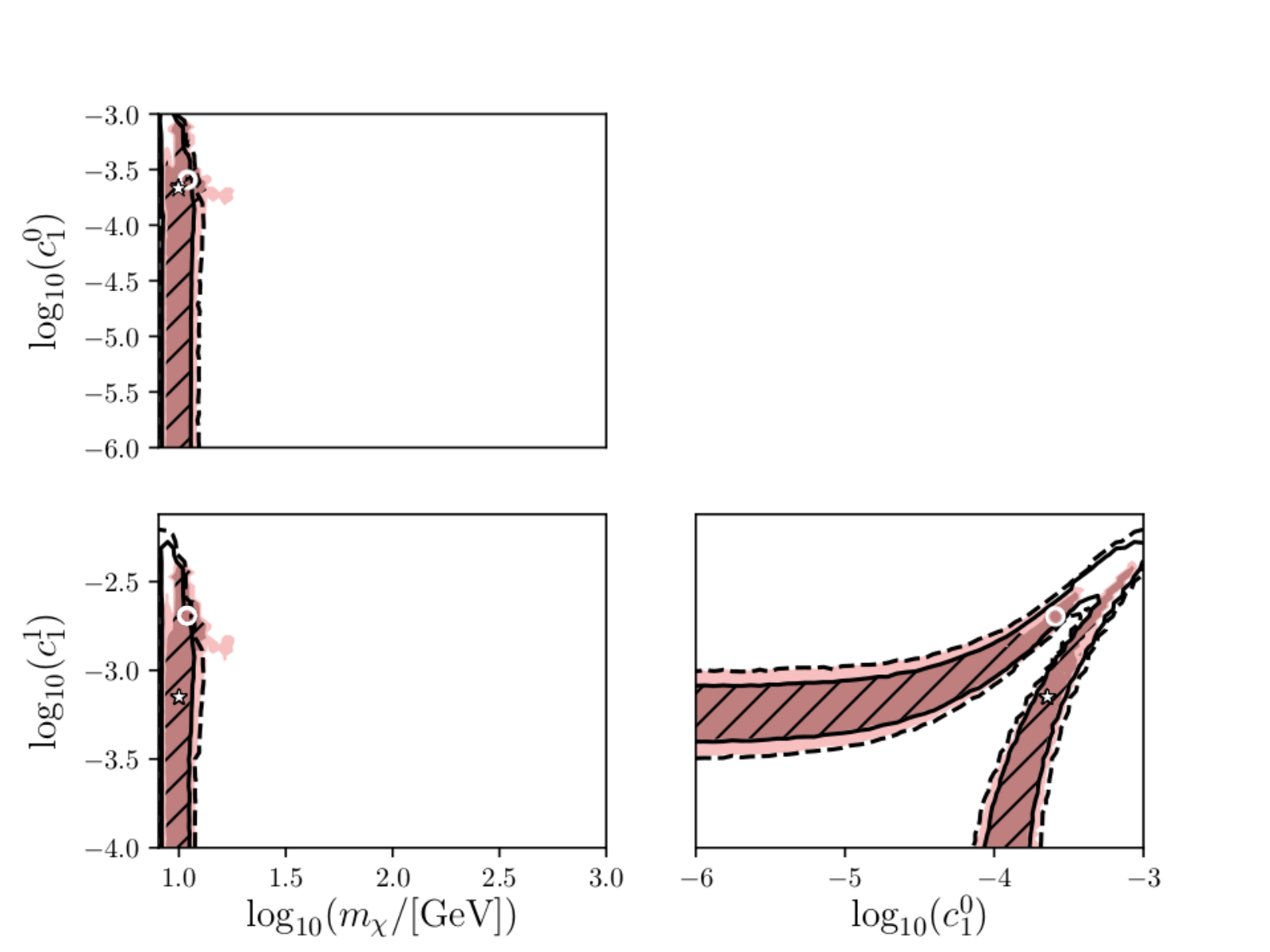}
\end{center}
\caption{Reconstruction of DM parameters in $(\mdm,\,c_1^0,\,c_{1}^1)$. The best-fit point and the $1\,\sigma$ (68\% C.L.) and $2\,\sigma$ (95\% C.L.) regions are shown for the results obtained with \acro\ (white star and shaded areas) and the full physics code (white ring and solid and dashed black lines).}
\label{fig:comparison_o1_01}
\end{figure}

We show in Figure\,\ref{fig:comparison_o1_01} the results of a three-dimensional scan $(\mdm,\,c_1^0,\,c_1^1)$ for a benchmark point that exhibits a large degree of fine-tuning. When comparing to the result using the full calculation, we can observe that our parametrisation method recovers the correct shape of the reconstructed areas, including the region where the negative interference takes place. As in previous examples, the time employed by our method was considerably shorter. 
Notice that the best fit point in our reconstruction (white star) does not coincide with that of the best-fit point (white ring) from the physics calculation. However, the best-fit point calculated by \acro $\,$is well within the $1\sigma$ contour.

Having proved that this prescription treats cancellations accurately, \acro's default setting is to produce output from a series of polynomials as described in equation (\ref{eq:param2}).

\subsection{Generalised DM haloes}
\label{sec:haloes}

The SHM is the canonical choice used to present the results from direct detection experiments. It is particularly convenient because the velocity integral can be solved analytically, which greatly reduces computing time. However, The SHM is in conflict with numerical simulations, which produce anisotropic DM halos with speed distributions that deviate systematically from the Maxwell-Boltzmann one \cite{Hansen:2005yj,Vogelsberger:2008qb,Kuhlen:2009vh,Ludlow:2011cs,
Mao:2012hf,Kuhlen:2013tra}. 
Direct detection experiments are very sensitive to the halo parameters \cite{Belli:2002yt,Pato:2010zk,Green:2011bv,Cerdeno:2016znc,Bozorgnia:2017brl}, especially when probing low-mass DM candidates \cite{Cerdeno:2016znc}. For example, changes in the velocity distribution function can significantly alter the population of DM particles with enough speed to produce recoils above the experimental threshold.

Our surrogate model can be easily adjusted to a general velocity distribution function. In fact, it is for general haloes that this method is more advantageous: whereas the full calculation relies on numerically solving the velocity integral, in our method, this only has to be done in the training phase.

As a final test of our method, we have applied our reconstruction routine to the same example as in the previous subsections, but considering a generalised DM halo, defined by a following velocity distribution function which differs from the SHM in a high-velocity tail 
\cite{Vogelsberger:2008qb,Ling:2009eh,Fairbairn:2008gz,Kuhlen:2009vh}. An additional parameter $k$ controls the deviations \cite{Lisanti:2010qx} as
\begin{equation}
	f(v) = N_k^{-1} \left[ e^{-v^2/k v_0^2} - e^{-v_{esc}^2/k v_0^2} \right]^k
	\Theta(v_{esc}-v),
	\label{eqn:F(v)}
\end{equation}
where $N_k=v_0^3 e^{-y_e^2}\int_{0}^{y_e}$ $dy~y^2(e^{-(y^2-y_e^2)/k}-1)^k$ 
and $y_e=v_{esc}/v_0$, and the SHM is recovered for $k=0$.
We have considered variations in the halo parameters as $v_{esc}\in[478, 610]$~km s$^{-1}$,  $v_0\in [170, 290]$~km s$^{-1}$, and $k\in[0.5, 3.5]$, which are included in our scan as nuisance parameters. 
The local DM density is also subject to observational uncertainties, and we have considered here a range $\rho_0\in [0.2, 0.6]$~GeV~cm$^{-3}$ \cite{Catena:2009mf,Salucci:2010qr,Pato:2010yq,Iocco:2011jz}. 
These ranges are consistent (although broader) than those obtained in  recent analysis of N-body simulations that include the effect of baryons Ref.\cite{Calore:2015oya,Bozorgnia:2016ogo}.

\begin{figure}
\begin{center} 
\includegraphics[scale=0.55, bb=0 0 460.8 345.6]{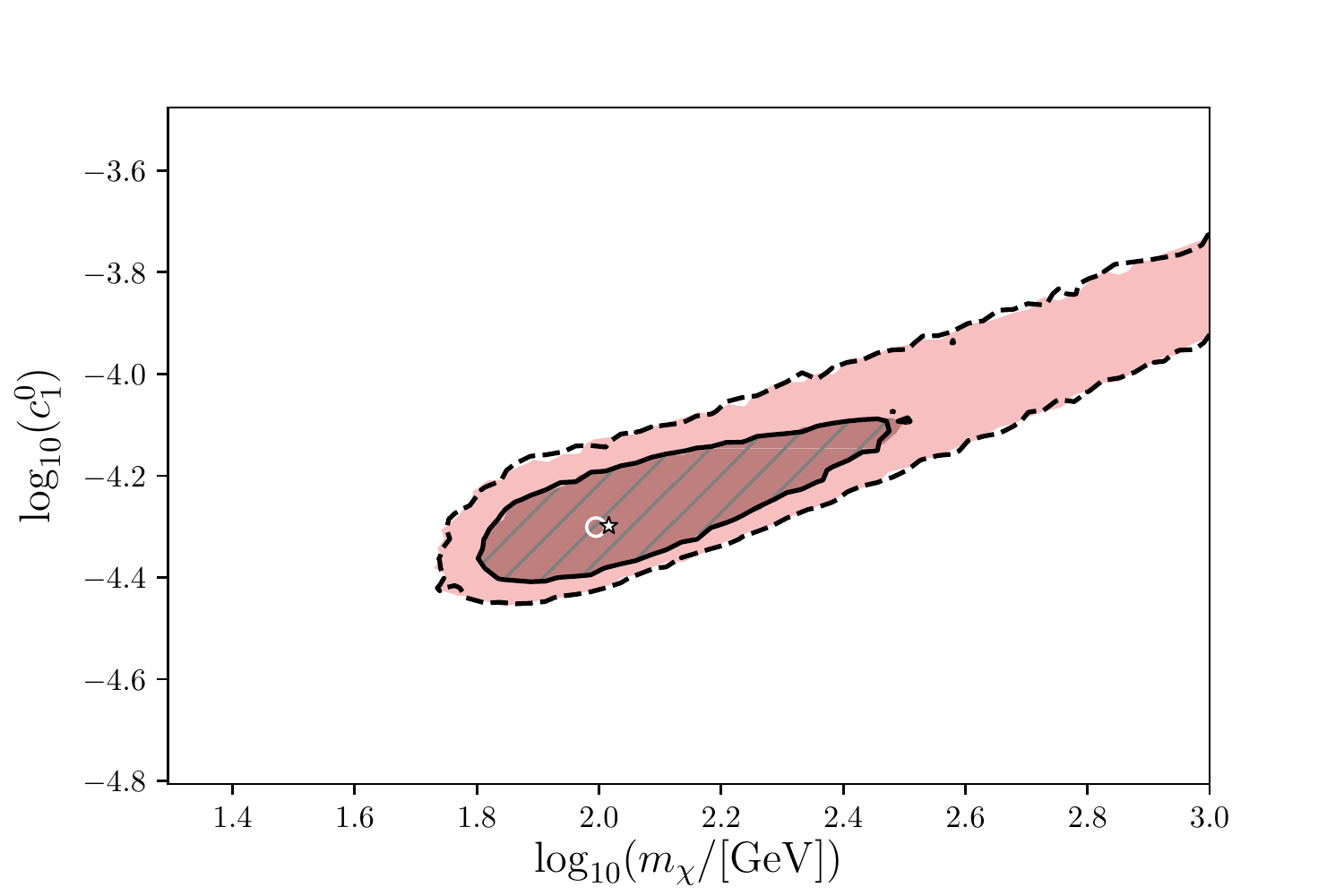}
\end{center}
\caption{Reconstruction of DM parameters in the $(\mdm,\,c_1^0)$ plane when astrophysical uncertainties are included. The best-fit point and the $1\,\sigma$ (68\% C.L.) and $2\,\sigma$ (95\% C.L.) regions are shown for the results obtained with \acro\ (white star and shaded areas) and the full physics code (white ring and solid and dashed black lines).}
\label{fig:astro}
\end{figure}

Figure\,\ref{fig:astro} shows the resulting reconstruction of DM parameters in this generalised halo. As expected, the $1\,\sigma$ (68\% C.L.) and $2\,\sigma$ (95\% C.L.) regions are wider as a consequence of astrophysical uncertainties. As in the previous examples, we observe no difference between the results obtained with \acro\ and those obtained with the full physics code.
We therefore conclude that our surrogate model is fast and reliable, and easily applicable to generalised DM haloes.

Finally, \acro\ can also incorporate velocity distribution functions which are defined as a function of the velocity, but not necessarily given by an analytical formula.
This is useful to interpret results from numerical simulations (see e.g., Refs.\,\cite{Bozorgnia:2016ogo,Bozorgnia:2017brl}).

\section{Simplified dark matter models}
\label{sec:simplified}
\setcounter{equation}{0}

Having proved the reliability and speed of \acro\ in the previous
section, we will now exploit this tool to illustrate what we consider as a major application of our method: DM parameter reconstruction in a multi-dimensional parameter space and model comparison in the presence of future data in direct detection experiments. As in the examples of the previous sections, we will employ MultiNest to efficiently sample the parameter space. When employing Bayesian inference methods, the reader should be aware that the posterior distribution functions (pdfs) can be sensitive to the choice of priors \cite{deAustri:2006jwj,Bertone:2011nj,Feroz:2011bj}. In our analysis below, we will show the profile likelihood, which  is usually more sensitive to small fine-tuned regions with large likelihood (while the integration implemented for the pdf accounts for volume effects). We should point out that \acro\ should be useful in both a frequentist or Bayesian approach. 

In this section we consider a set of simplified DM models, in which a dark matter particle and a new mediator are added to the SM Lagrangian. The EFT operators for direct detection can be recovered in the non-relativistic limit \cite{Dent:2015zpa}. In this section, simplified models will be labelled according to the nature of the DM and mediator as follows: SS (scalar DM with scalar or pseudoscalar mediator), SV (scalar DM with vector mediator),  FS (fermion DM with scalar or pseudoscalar mediator), FV (fermion DM with vector mediator)\footnote{We have not included examples with vector DM, although we have explicitly checked that they are successfully reproduced with \acro. The reason is that in these models, when the mediator is a scalar the operator space is the same as in the SS case, and when the mediator is a vector, the parameter space contains up to eight operators. This is difficult to constrain using just the three experiments of Table 1: only upper bounds are obtained on the EFT operators and the results are therefore not very enlightening. }.

Assuming that the couplings in the Lagrangian are of the same order, there is a subset of EFT operators that give the leading contributions to the DM-nucleus scattering rate for each simplified DM model, as studied in Refs.\,\cite{Dent:2015zpa,Baum:2017kfa}. This allows us to design strategies that can lead to the discrimination of DM models \cite{Schneck:2015vtw,Gluscevic:2015sqa,Witte:2016ydc,Catena:2017wzu,Kahlhoefer:2017ddj,Baum:2017kfa}
There are public codes, such as the one presented in Ref.~\cite{Bishara:2017nnn} that can be used to obtain this matching. 
We will use these subsets of operators to reconstruct DM parameters from a hypothetical signal, thus testing the validity of each individual scenario. For concreteness, the relation between models and operators is as follows: SS $\left\{\op{1},\, \op{10}\right\}$, SV $\left\{\op{1},\,\op{7}\right\}$, FS $\left\{\op{1},\,\op{6},\,\op{10},\,\op{11}\right\}$, FV $\left\{\op{1},\,\op{4},\,\op{7},\,\op{8},\,\op{9}\right\}$.
It should be emphasized at this point that among these operators, $\op{5}$, $\op{6}$, $\op{9}$, $\op{10}$ and $\op{11}$ have a non-trivial momentum dependence that leads to an unconventional spectrum (which vanishes for $E_R\to0$).

We will  consider a hypothetical future situation in which several direct detection experiments observe an excess in their data that can be attributed to DM particles. We will attempt to reconstruct the data within the context of different simplified models. As already mentioned in the Introduction, in general, a single experimental target is unable to unambiguously determine the DM couplings, thus we consider a signal in three targets, Ge, Xe, and Ar.

\begin{table}[t!]
	\begin{center}
	\begin{tabular}{|l |l | l|| l| l| l|}
	\hline
    Name&Model& DM Parameters &$N_{\rm Xe}$&$N_{\rm Ge}$&$N_{\rm Ar}$\\
    \hline
    BP1 & SS & 
    \begin{minipage}{3cm}
    \rule{0ex}{0.ex}\\
    $\mdm=10$ GeV \\ 
    $c_1=1\times10^{-4}$\\  
    $c_{10}=5$
    \vspace*{0.8ex}
    \end{minipage}
     &$93$&$10$&$50$\\
\hline
BP2 & SS & 
        \begin{minipage}{3cm}
        \rule{0ex}{0.ex}\\
        $\mdm=100$ GeV\\  
        $c_1=3\times10^{-5}$\\  
        $c_{10}=5\times 10 ^{-1}$
        \vspace*{0.8ex}
       \end{minipage}&$206$&$2$&$30$\\
\hline
    BP3 & FS &
        \begin{minipage}{3cm}
        \rule{0ex}{0.ex}\\
        $\mdm=30$ GeV\\  
        $c_1=0.0$\\  
        $c_6=60$\\
        $c_{10}=0.0$\\
        $c_{11}=0.0$
        \vspace*{0.8ex}
       \end{minipage}&$256$&$1$&$0$\\
	\hline
	\end{tabular}
		\end{center}
    \caption{Benchmark points considered in this paper. They all satisfy current experimental constraints from direct detection experiments and are within the reach of next generation detectors. For reference, we indicate the total number of DM events expected in each of the experimental configurations of Table\,\ref{tab:experiments}.}
    \label{tab:BPs}
\end{table}

We have selected a number of benchmark points, shown in Table~\ref{tab:BPs}, all of which satisfy the current experimental bounds from direct detection experiments. 
We include one example with a low-mass DM particle (BP1) and another one with a heavier candidate (BP2), since they give rise to different issues in the parameter reconstruction. We have also chosen a point motivated by Pseudoscalar-mediated DM \cite{Gresham:2014vja,Chang:2010yk} (BP3).
We have assumed universal couplings of the DM to quarks, which leads to a specific relation between the isoscalar and isovector components of the DM-nucleus coupling (see, for example, Ref.\,\cite{Baum:2017kfa}), thereby effectively reducing the dimensionality of the parameter space.
For each benchmark point, we have generated mock data for the experimental setups of Table\,\ref{tab:BPs}. Then, using this data, we have attempted the reconstruction of DM parameters (mass and couplings) for each simplified DM model (SS, SV, FS, FV, VS, VV) using \acro\, linked with MultiNest. In all cases, we have computed the reconstruction corresponding to each individual target, as well as the one resulting from the combination of data from the three targets. For clarity, the plots showing the resulting profile likelihood in the multi-dimensional parameter space are shown in Appendix~\ref{sec:appendix}, and in the rest of this section we will only show the DM spectra corresponding to the best-fit points in each model.

In this case study we have not computed the DM relic abundance, as it  collider constraints. The latter might be particularly relevant to some specific EFT operator. A recent analysis \cite{Bertone:2017adx} has explored the combination of direct detection experiments with collider constraints in order to identify the simplified model. These bounds can be easily implemented in the reconstruction algorithm, although they lie beyond the scope of our analysis.

\subsection{BP1 (light DM):}

Our first benchmark point, with $\mdm=10$~GeV, is an example of low-mass DM candidate.  When doing the matching for non-relativistic operators for scalar mediated DM one finds only $\op{1}$ and $\op{10}$ responses \cite{Dent:2015zpa,Catena:2017wzu}. Interestingly these two operators have a different momentum dependence. However, due to the small DM mass, the characteristic shape of $\op{10}$ can be mistaken for a typical exponential behaviour unless the experimental threshold is very low. This is a challenge for parameter reconstruction that can be alleviated through the use of multiple targets.

\begin{figure}
\hspace*{-2cm}
\includegraphics[width=19cm, bb=0 0 1152 576]{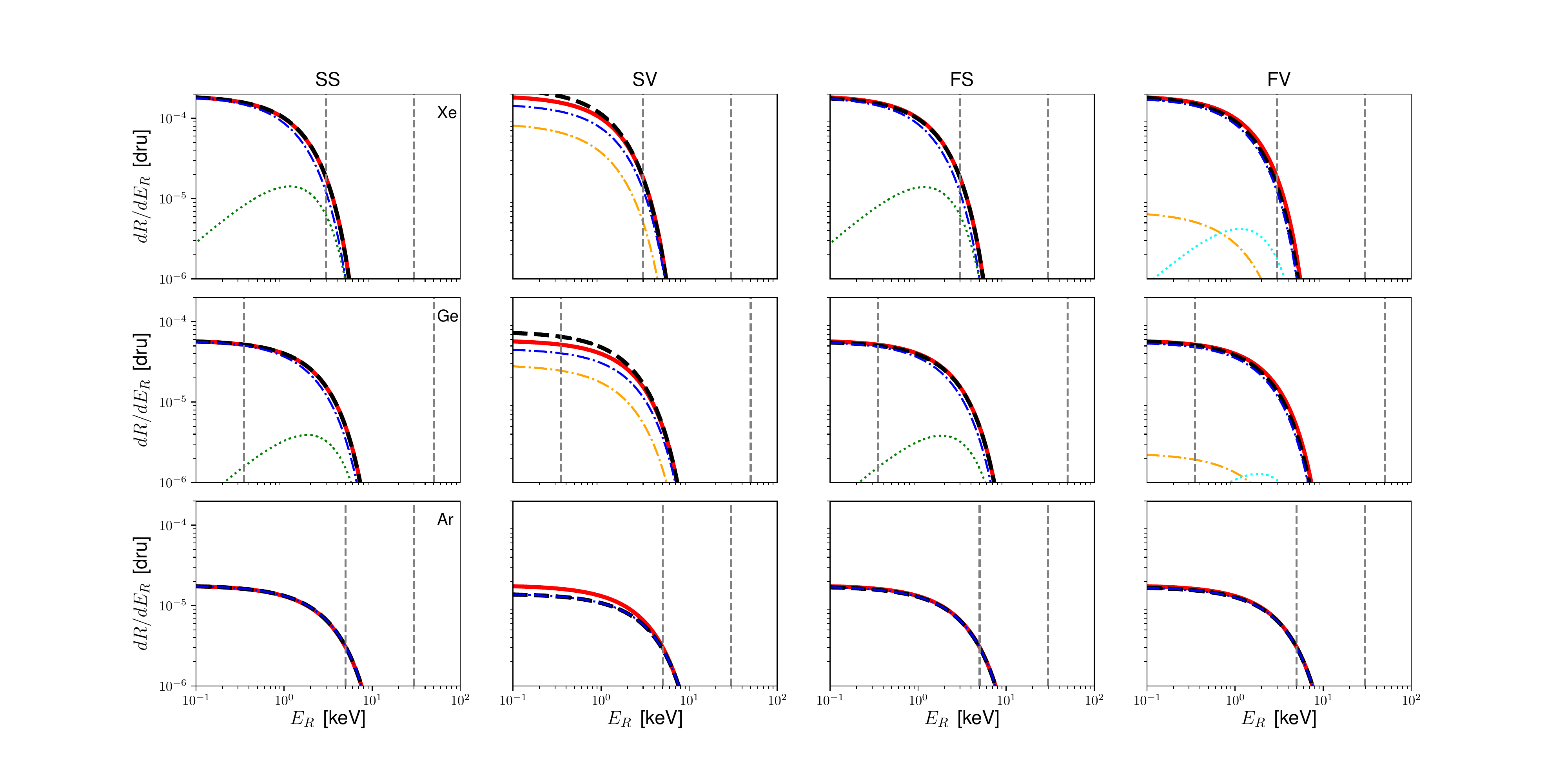}
\hspace*{0.2cm}{\small\begin{tabular}{|l| l |l| l |l| l |l|}
\hline
\begin{minipage}{3cm}
    \rule{0ex}{1ex}\\
    $\mdm=10.3$ GeV \\ 
    \textcolor{blue}{$c_1=1.01\times10^{-4}$}\\  
    \textcolor{ForestGreen}{$c_{10}=4.89$} \\
    \rule{0ex}{1ex}
    \end{minipage} 
& &
 \begin{minipage}{3cm}
    \rule{0ex}{1ex}\\
    $\mdm=11.2$ GeV\\ 
    \textcolor{blue}{$c_1=9.23\times10^{-5}$}\\  
    \textcolor{BurntOrange}{$c_{7}=69.4$} \\
    \rule{0ex}{1ex}
    \end{minipage}     
& &
    \begin{minipage}{3cm}
    \rule{0ex}{1ex}\\
    $\mdm=10.3$ GeV\\ 
    \textcolor{blue}{$c_1=9.93\times10^{-5}$}\\  
    \textcolor{BurntOrange}{$c_{6}=24.1$} \\
    \textcolor{ForestGreen}{$c_{10}=4.80$} \\
   	\textcolor{Plum}{$c_{11}=1.10\times 10^{-4}$} \\
    \rule{0ex}{1ex}
    \end{minipage}
& &
\begin{minipage}{3cm}
    \rule{0ex}{1ex}\\
    $\mdm=10.5$ GeV\\ 
    \textcolor{blue}{$c_1=9.85\times10^{-5}$}\\  
    \textcolor{cyan}{$c_{4}=4.30\times 10^{-3}$} \\
    \textcolor{BurntOrange}{$c_{7}=18.7$} \\
   	\textcolor{ForestGreen}{$c_{8}=1.52\times 10^{-2}$} \\
   	\textcolor{cyan}{$c_{9}=-1.66$} \\
    \rule{0ex}{1ex}
    \end{minipage} 
\\
 \cellcolor{lightgray}$\log\mathcal{L}({SS}) = -29.9$&&
$\log\mathcal{L}({SV})=-30.3$&&
$\log\mathcal{L}({FS})=-29.9$&&
$\log\mathcal{L}({FV})=-33.9$
\\
\hline
\end{tabular}} \\
\caption{Reconstruction of parameters for BP1: Differential rate as a function of the recoil energy corresponding to the best fit point in each simplified models (columns) and for each of the experimental targets (rows).
The thick black line corresponds to the full differential rate obtained from the best fit point (after combination of data from the three targets) in each of the simplified models. For reference, the thick red line shows the differential rate corresponding to the benchmark point. The thin dotted (dot-dashed) lines represent the individual contributions from momentum dependent (independent) operators. The vertical dashed lines delimit the energy range explored for each target. The table indicates the parameters for the best fit points in each case (using the same colour code as the lines in the plots), and the value for its log-likelihood calculated using eq.~\ref{eqn:likelihood}. Gray shading is used to denote the true model (SS in this case).}
\label{fig:spect-bp1}
\end{figure}

In Figure~\ref{fig:spect-bp1}, we show the DM differential rate obtained for the best fit points in each simplified model and target (the different columns represent, from left to right, SS, SV, FS, and FV, and the different rows represent, from top to bottom, Ge, Xe, and Ar). The vertical grey dashed lines represent the energy range used in the fit for each target. 
In each plot, the red line corresponds to the differential rate predicted by the benchmark point (BP1), and the thick, dashed, black line is the differential rate obtained for the best-fit point (combining the data of the three targets). The individual contributions from EFT operators are shown by means of a dot-dashed line (for operators with a canonical momentum dependence) and dotted line (for operators with an extra momentum-dependence). 
The table below the plot indicates the parameters for the best-fit point in each simplified model, using the same colour code as the figure.

The full reconstruction can be found in Figures~\ref{fig:bp1ss} (SS), ~\ref{fig:bp1sv} (SV), ~\ref{fig:bp1fs} (FS) ~\ref{fig:bp1fv} (FV) of Appendix~\ref{sec:appendix}, where the profile likelihood in the DM parameters is represented. In these plots we have indicated the $2\,\sigma$ (95\% C.L.) contours obtained with each individual target, germanium (blue), xenon (green), and argon (orange). The combined results are shown by means of a shaded area and black dashed and solid lines for the $1\,\sigma$ (68\% C.L.) and $2\,\sigma$ (95\% C.L.) contours.
As we can observe in these figures, the DM mass is very well-reconstructed around the nominal value. Given the small DM mass, the end-point of the DM spectrum falls within the energy range explored in the three experiments. Notice that this argument is independent of the effective operator (the right mass is obtained in all scenarios), and therefore it does not help in discriminating the different models. As expected, a single experiment is unable to unambiguously determine the DM parameters. This can be seen, e.g., in Figures~\ref{fig:bp1ss} (SS), where the lines corresponding to Ge, Xe, and Ar, fail to produce closed contours in the $(c_1,\,c_{10})$ plane. This same effect happens for all other models, Figs.~\ref{fig:bp1sv} (SV), ~\ref{fig:bp1fs} (FS), and ~\ref{fig:bp1fv} (FV).

For the low masses that we are considering, the responses of operators $\op{1}$ and $\op{10}$ are very similar in both germanium and xenon.  Thus, these two targets are unable to resolve the degeneracy and in principle only are able to place an upper bound on the corresponding couplings.
We can observe this effect in Figure \ref{fig:bp1ss}, the $\{\op{1}, \op{10}\}$ plane, where the Germanium contours are not closed.
Interestingly, argon is  insensitive to the spin-dependent interaction $\op{10}$, which makes it ideal to resolve the ambiguity. This leads to closed contours in both SS (Fig.~\ref{fig:bp1ss}) and FS  (Fig.~\ref{fig:bp1fs}).

In all the four models, the reconstruction favours a leading contribution from operator $\op{1}$, consistent with the original benchmark point. The small contribution from a momentum-dependent operator is either attributed to $\op{10}$, (in SS and FS) or $\op{9}$ (in FV).  
In order to quantify and compare the goodness of the resulting fits, we have computed the log-likelihood of each best-fit point, as given by eq.~(\ref{eqn:likelihood}). 
We can observe that a relatively good fit is obtained in all four scenarios, with a slight preference for the right model SS and also FS (since it has the same operators).\footnote{
Notice that in this kind of analysis, it is customary to compare hypothesis by means of the Bayesian evidence, however, we are here dealing with models of different dimensionality and we would observe the rather trivial result that models with more free parameters are favoured. As explained for example in Ref.\,\cite{Rogers:2016jrx}, one could start by calculating the evidence for 2D slices of the parameter space and thus identify the most likely set of parameters before moving to larger dimensions.}

Finally, as a consistency check, we have compared the binned DM spectrum for the best fit points obtained with \acro\ and with the full physics code. We have observed that the number of DM events per bin obtained with the surrogate model and the real one differ by less than 1\% when fourth order polynomials are employed for models SS, SV, and FS. We have found that model FV requires a fifth order polynomial to attain the same degree of precision.

\subsection{BP2 (Heavy DM)}

\begin{figure}
\hspace*{-2cm}
\includegraphics[width=19cm, bb=0 0 1152 576]{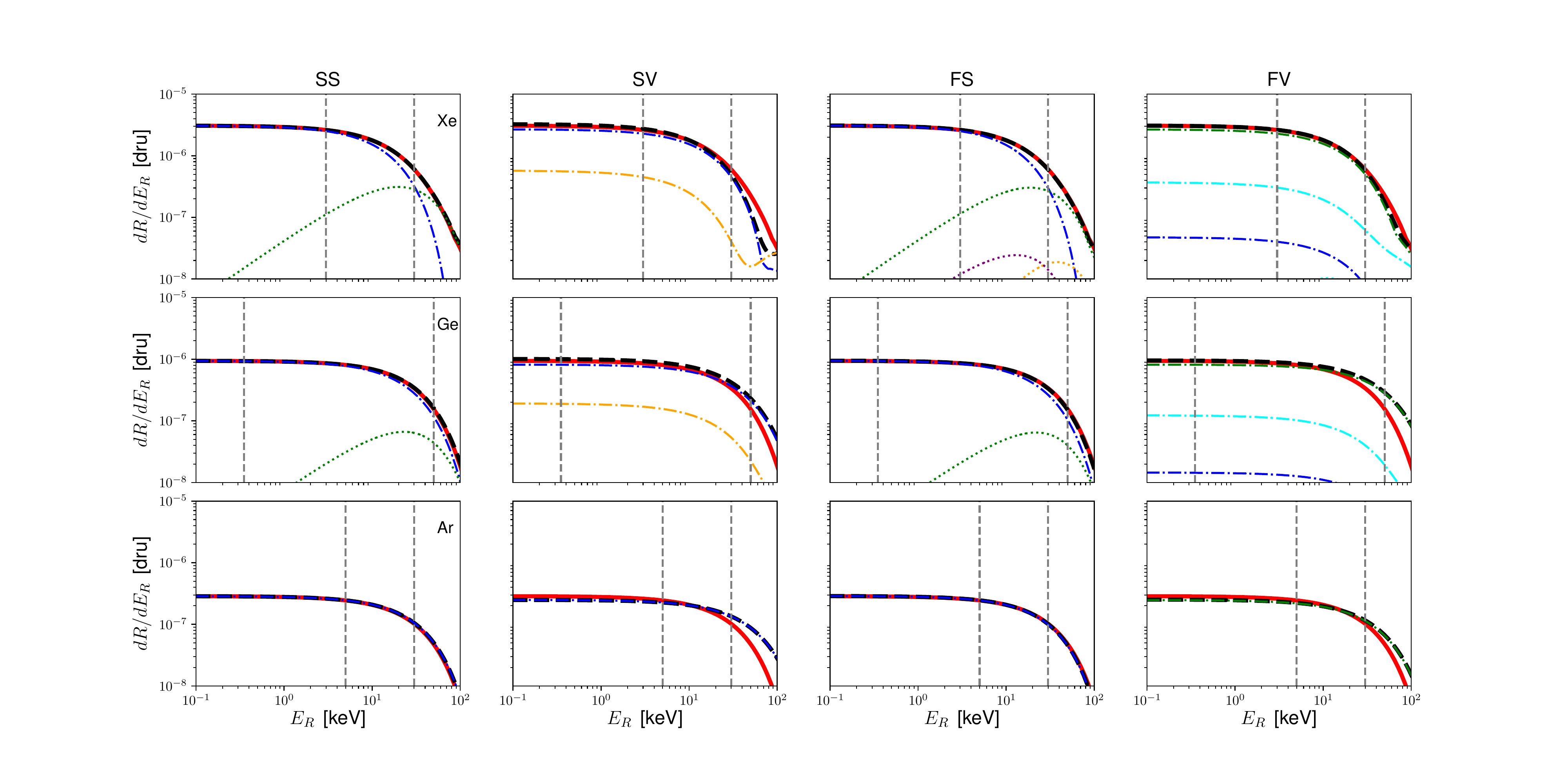}
\hspace*{0.2cm}{\small\begin{tabular}{|l| l |l| l |l| l |l|}
\hline
 \begin{minipage}{3cm}
    \rule{0ex}{1ex}\\
    $\mdm=105$ GeV\\ 
    \textcolor{blue}{$c_1=4.11\times 10^{-5}$}\\  
    \textcolor{ForestGreen}{$c_{10}=0.595$} \\
    \rule{0ex}{1ex}
    \end{minipage} 
&&
 \begin{minipage}{3cm}
    \rule{0ex}{1ex}\\
    $\mdm=989$ GeV\\ 
    \textcolor{blue}{$c_1=3.79\times10^{-4}$}\\  
    \textcolor{BurntOrange}{$c_{7}=161$} \\
    \rule{0ex}{1ex}
    \end{minipage}
&&
\begin{minipage}{3cm}
    \rule{0ex}{1ex}\\
    $\mdm=95.7$ GeV\\ 
    \textcolor{blue}{$c_1=3.90\times10^{-5}$}\\  
    \textcolor{BurntOrange}{$c_{6}=3.22$} \\
    \textcolor{ForestGreen}{$c_{10}=0.570$} \\
   	\textcolor{Plum}{$c_{11}=1.85\times 10^{-4}$} \\
    \rule{0ex}{1ex}
    \end{minipage}
&&
\begin{minipage}{3cm}
    \rule{0ex}{1ex}\\
    $\mdm=1403$ GeV \\ 
    \textcolor{blue}{$c_1=1.81\times10^{-5}$}\\  
    \textcolor{cyan}{$c_{4}=5.14\times 10^{-2}$} \\
    \textcolor{BurntOrange}{$c_{7}=1.69$} \\
   	\textcolor{ForestGreen}{$c_{8}=3.04\times 10^{-1}$} \\
   	\textcolor{cyan}{$c_{9}=0.303$} \\
    \rule{0ex}{1ex}
    \end{minipage}
\\ 
 \cellcolor{lightgray}$\log\mathcal{L}({SS}) = -85.2$&&
$\log\mathcal{L}({SV})=-86.0$&&
$\log\mathcal{L}({FS})=-85.2$&&
$\log\mathcal{L}({FV})=-85.4$
\\
\hline
\end{tabular}}\\
\caption{The same as in Fig.\,\ref{fig:spect-bp1}, but for benchmark point BP2.}
\label{fig:spect-bp2}
\end{figure}

We now turn our attention to a larger value of the DM mass. Benchmark point BP2 features a 100~GeV particle. This implies that the resulting spectrum is flatter and displaced towards larger values of the recoil energy.  
It should be noted that with the configurations chosen in Table\,\ref{tab:experiments}, only the xenon and argon targets would be sensitive to this signal. Since we have assumed a smaller exposure for germanium, the expected number of events for this target is merely $N_{\rm Ge}=1$, which only leads an upper bound in the corresponding couplings.
The spectra for the best-fit points are shown in Fig.~\ref{fig:spect-bp2}.

Due to the heavier DM mass, the endpoint of the recoil spectrum lies beyond the energy window of all three targets, which makes its reconstruction more difficult. Moreover, the (small) contribution from the momentum-dependent operator $\op{10}$ flattens out the spectrum at large energies. This is properly identified in models SS and FS (for which the resulting value of $c_{10}$ is comparable to that of the original benchmark point), resulting also in a very good reconstruction of the DM  mass. However, in models SV and FV for which there are no momentum-dependent operators, the best fit is obtained for a much larger value of the DM mass (in model SV the best fit is actually towards the boundary of the reconstructed area with $\mdm\sim1000$~GeV), so as to compensate for the flatter spectrum.

The full profile likelihood in the whole parameter space can be found in  Figures~\ref{fig:bp2ss} (SS), ~\ref{fig:bp2sv} (SV), ~\ref{fig:bp2fs} (FS), and ~\ref{fig:bp2fv} (FV) of Appendix~\ref{sec:appendix}. 
As we see in these plots, the reconstructed DM mass has a large uncertainty, and even if the right value is obtained for the best-fit point (model SS), the $2\,\sigma$ (95\% C.L.) region is unbounded from above. This effect is well known, and is exacerbated by astrophysical uncertainties in the DM escape velocity \cite{McCabe:2010zh,Pato:2010zk,Kavanagh:2014rya}. 
As we can see in the table below Figure~\ref{fig:spect-bp2}, the goodness of the reconstruction is very similar for models SS and FS. This is not surprising, since once more the response in both cases is dominated by the same set of operators. These two models are difficult to disentangle using direct detection alone, but as recent analyses point out, a combination with LHC data could shed light onto the nature of the DM and the mediator \cite{Bertone:2017adx}.
The goodness of the fit for models SV and FV is not much worse, but the reconstructed areas are extremely degenerate. For example in model FV there is a complete degeneracy between operators $\op{1}$ and $\op{8}$.

\subsection{Momentum-dependent DM (BP3)}

Finally, we have selected an example based on fermion DM with a pure pseudo-scalar mediator (model FS with only operator $\op{6}$), since this gives rise to a very characteristic spectrum which vanishes at small recoil energies. Notice that given our choice of parameters for benchmark point BP3 of Table \ref{tab:BPs}, only xenon sees a relevant number of DM events. However, the data from the other targets is still useful to set up upper bounds on specific operators.

\begin{figure}
\hspace*{-2cm}
\includegraphics[width=19cm, bb=0 0 1152 576]{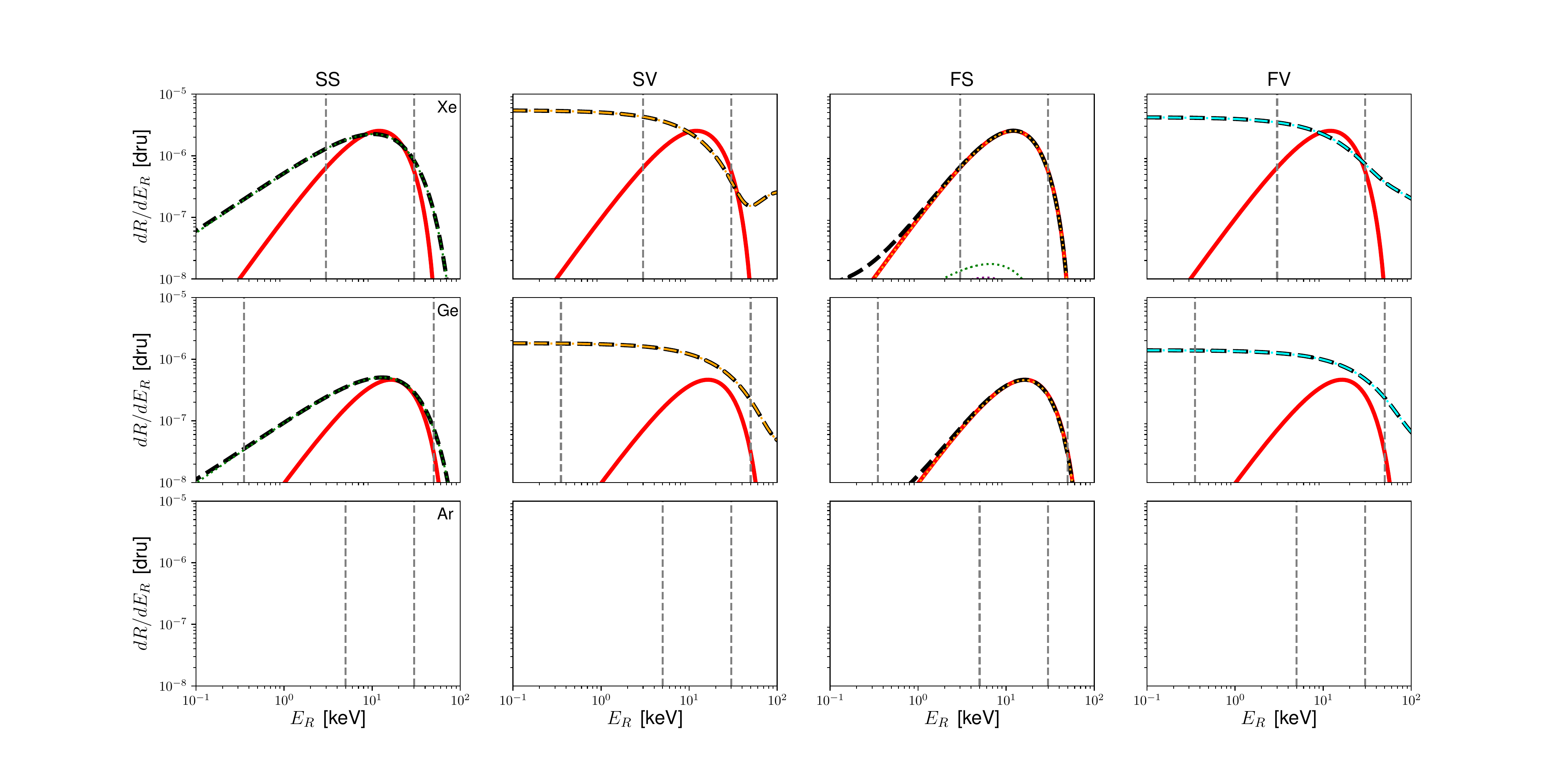}
\hspace*{0.2cm}{\small\begin{tabular}{|l| l |l| l |l| l |l|}
\hline
\begin{minipage}{3cm}
   \rule{0ex}{1ex}\\
    $\mdm=47.0$ GeV\\ 
    \textcolor{blue}{$c_1=1.06\times 10^{-6}$}\\  
    \textcolor{ForestGreen}{$c_{10}=1.39$} \\
    \rule{0ex}{1ex}
    \end{minipage} 
&&
\begin{minipage}{3cm}
    \rule{0ex}{1ex}\\
    $\mdm=9920$ GeV\\ 
    \textcolor{blue}{$c_1=2.48\times10^{-6}$}\\  
    \textcolor{BurntOrange}{$c_{7}=4.91\times10^{2}$} \\
    \rule{0ex}{1ex}
    \end{minipage}
&&
\begin{minipage}{3cm}
    \rule{0ex}{1ex}\\
    $\mdm=30.9$ GeV\\ 
    \textcolor{blue}{$c_1=1.00\times10^{-6}$}\\  
    \textcolor{BurntOrange}{$c_{6}=59,8$} \\
    \textcolor{ForestGreen}{$c_{10}=0.126$} \\
   	\textcolor{Plum}{$c_{11}=1.02\times 10^{-4}$} \\
    \rule{0ex}{1ex}
    \end{minipage}
&&
\begin{minipage}{3cm}
    \rule{0ex}{1ex}\\
    $\mdm=9966$ GeV\\ 
    \textcolor{blue}{$c_1=2.35\times10^{-6}$}\\  
    \textcolor{cyan}{$c_{4}=0.469$} \\
    \textcolor{BurntOrange}{$c_{7}=1.40$} \\
   	\textcolor{ForestGreen}{$c_{8}=1.21\times 10^{-2}$} \\
   	\textcolor{cyan}{$c_{9}=1.20\times10^{-2}$} \\
    \rule{0ex}{1ex}
    \end{minipage}
\\
$\log\mathcal{L}({SS}) = -62.6$&&
$\log\mathcal{L}({SV})=-103$&&
\cellcolor{lightgray}$\log\mathcal{L}({FS})=-58.9$&&
$\log\mathcal{L}({FV})=-84.4$
\\
\hline
\end{tabular}}\\
\caption{The same as in Fig.\,\ref{fig:spect-bp1}, but for benchmark point BP3. No events are expected for argon in the last row of plots.}
\label{fig:spect-bp3}
\end{figure}

Figure \ref{fig:spect-bp3} shows the differential DM rate corresponding to the best fit points in each model. We can now observe that only the right scenario (FS) produces a good fit to the signal. The reason is that  $\op{6}$ is the only operator of the set considered here that is $q^2$ dependent.  In contrast, $\op{10}$ only depends on $q$ and thus leads to a different shape.
As the table below Figure \ref{fig:spect-bp3} shows, the statistically preferred model coincides with the true model quite unequivocally.

The full parameter reconstruction can be found in Figures~\ref{fig:bp3ss} (for model SS), ~\ref{fig:bp3sv} (SV), ~\ref{fig:bp3fs} (FS), and ~\ref{fig:bp3fv} (FV) of Appendix~\ref{sec:appendix}. As expected, the best fit areas for the true model FS in Figure~\ref{fig:bp3fs} correctly identify the leading role of operator $\op{6}$. In model SS, the best fit is obtained for large $\op{10}$,
as it also has a non-trivial momentum dependence.
Finally, when models SV or FV are used in the reconstruction, there is a substantial tension between the areas obtained using only xenon data and those using the other targets. If only xenon data is considered, the best fit areas favour a leading contribution from operator $\op{1}$. However, this would be inconsistent with the non-observation of events in argon. When argon data is included, the best-fit area corresponds to large values of $\op{7}$ in model SV or $\op{4}$ in FV (to which argon is insensitive). This tension manifests as a much poorer fit, as observed in the table below Figure \ref{fig:spect-bp3}. 

As in previous examples, we have checked the accuracy of the surrogate model in the best fit points, obtaining a difference of less than a 1\% with the full physics computation.

\section{Conclusions and prospects}
\label{sec:conclusions}

In this article we have introduced \acro, a surrogate model to compute the binned DM spectrum in direct detection experiments. \acro\ substitutes the full physics computation (which in general involves up to three nested integrals) with a much faster parametrization in terms of ordinary polynomials of the DM mass and couplings. The surrogate model is initially trained for a given choice of parameters using the full calculation of the DM rate for a given set of direct detection experiments. The parametrization is then extracted using the {\sc PROFESSOR} tool.

We have validated our surrogate model using a range of examples that explore the reconstruction of DM parameters using mock-data in the multi-dimensional parameter space of effective field theories motivated from simplified DM models. We have identified and overcome two difficulties, corresponding to the case of low-mass dark matter and the interference between different operators. We have also checked that \acro\ can successfully incorporate a generic DM halo, as such obtained from N-body simulations, and include astrophysical uncertainties in the halo parameters. In a few selected benchmark points, we have compared our results with those of the full physics calculation, obtaining a perfect agreement and a runtime approximately two orders of magnitude smaller.

As a final test of the full potential of \acro, we have attempted the reconstruction of DM parameters in the context of a set of simplified models using three experimental setups, inspired by future DM detectors. We have considered the cases of scalar and fermion DM particles, with either scalar or vector mediators. In these models, the dimensionality of the parameter space (once astrophysical uncertainties are included as nuisance parameters) ranges from six to nine dimensions and include operators with non trivial momentum dependence. We have found that fourth order polynomials provide a good fit, with errors smaller than 1\%, except for the example with highest dimensionality, where fifth order polynomials were required. Using three experimental targets (Ge, Xe, and Ar), we have illustrated the advantage of target complementarity. Although in general, the right model cannot be fully determined due to the limitations of experimental data, analyses like this one can be used to assess the suitability of future experimental targets.

In conclusion, \acro\ is well suited to perform fast and accurate scans in a large number of dimensions. It is therefore ideal to explore the wide parameter space of effective field theory operators and could be used by experimental collaborations for a quick interpretation of their results. 
\acro\ can also be used in scans that require a large number of evaluations, such as in global scans of particle physics models.

\vspace*{1cm}
\noindent{\bf \large Acknowledgements}

We thank N. Bozorgnia, R. Catena, T. Jubb, M. Krauss, J. Men\'endez, and M. Peir\'o. This material is based upon work supported by the U.S. Department of Energy, Office of Science, Office of Advanced Scientific Computing Research, Scientific Discovery through Advanced Computing (SciDAC) program and the Science and Technology Facilities Council (STFC). 


\providecommand{\href}[2]{#2}\raggedright

\clearpage

\appendix
\section{Complementary plots}
\label{sec:appendix}

In this appendix we include the plots that correspond to the DM parameter reconstruction of simplified DM models in Section~\ref{sec:simplified}. We present here the resulting profile likelihood in the multidimensional parameter space. 

\subsection{BP 1}
\label{sec:bp1}

\begin{figure}[h!]
\begin{minipage}{7cm}
\includegraphics[scale=0.52, bb=0 0 446.4 432]{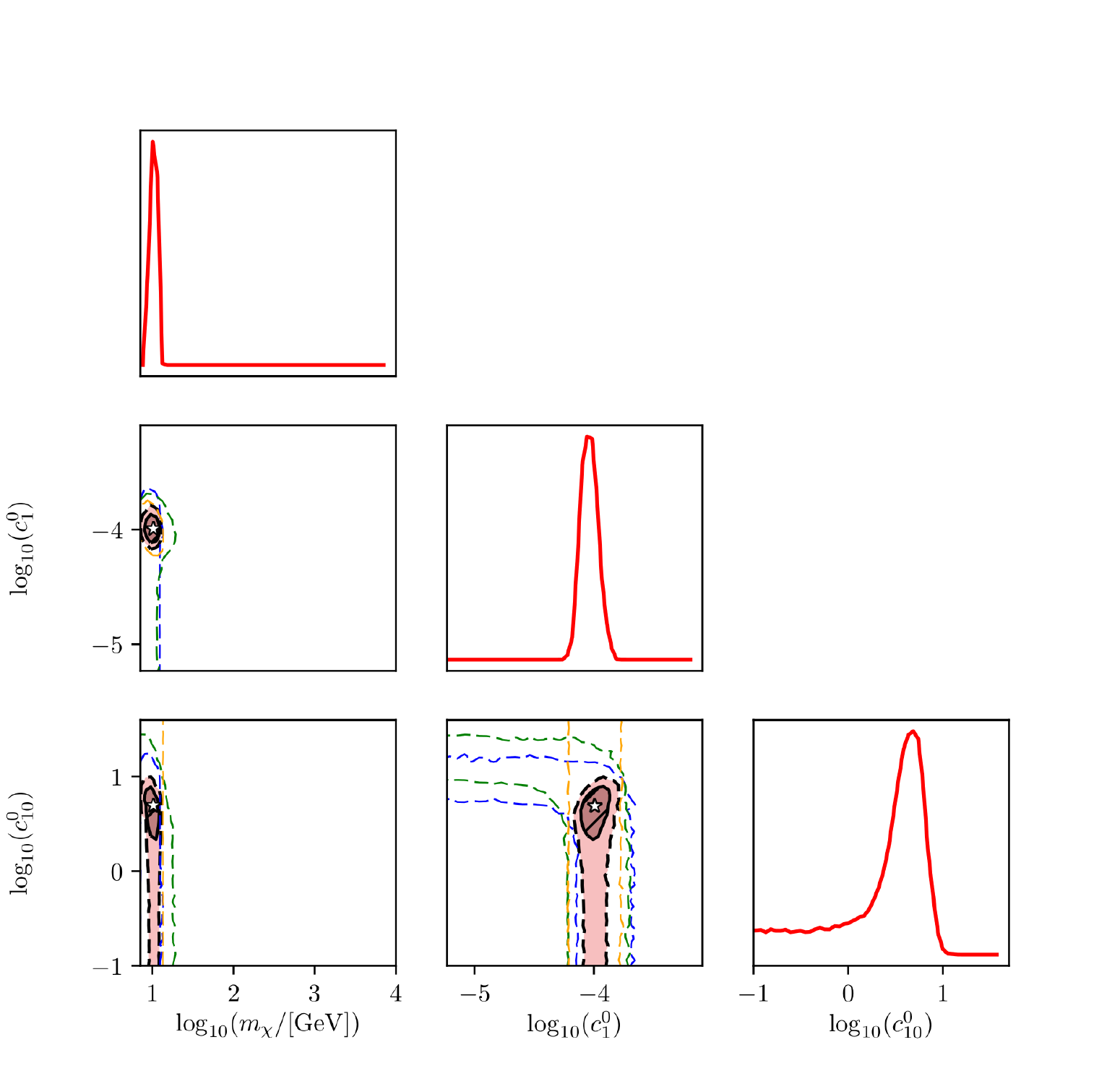}
\end{minipage}
\hspace*{1cm}\begin{minipage}{7cm}
\caption{Profile likelihood for the reconstruction of DM parameters for the simulated data of benchmark point BP1, using simplified model SS. Dashed blue, green, and orange lines correspond to the $2\,\sigma$ (95\% C.L.) contours obtained for individual targets of xenon, germanium, and argon. Black dashed and solid lines correspond to the contours obtained for the combination of the three targets. The best fit point is represented by an asterisk. For reference, the one-dimensional profile likelihoods are also shown.}
\label{fig:bp1ss}
\end{minipage}
\end{figure}

\begin{figure}[h!]
\begin{minipage}{7cm}
\includegraphics[scale=0.52, bb=0 0 446.4 432]{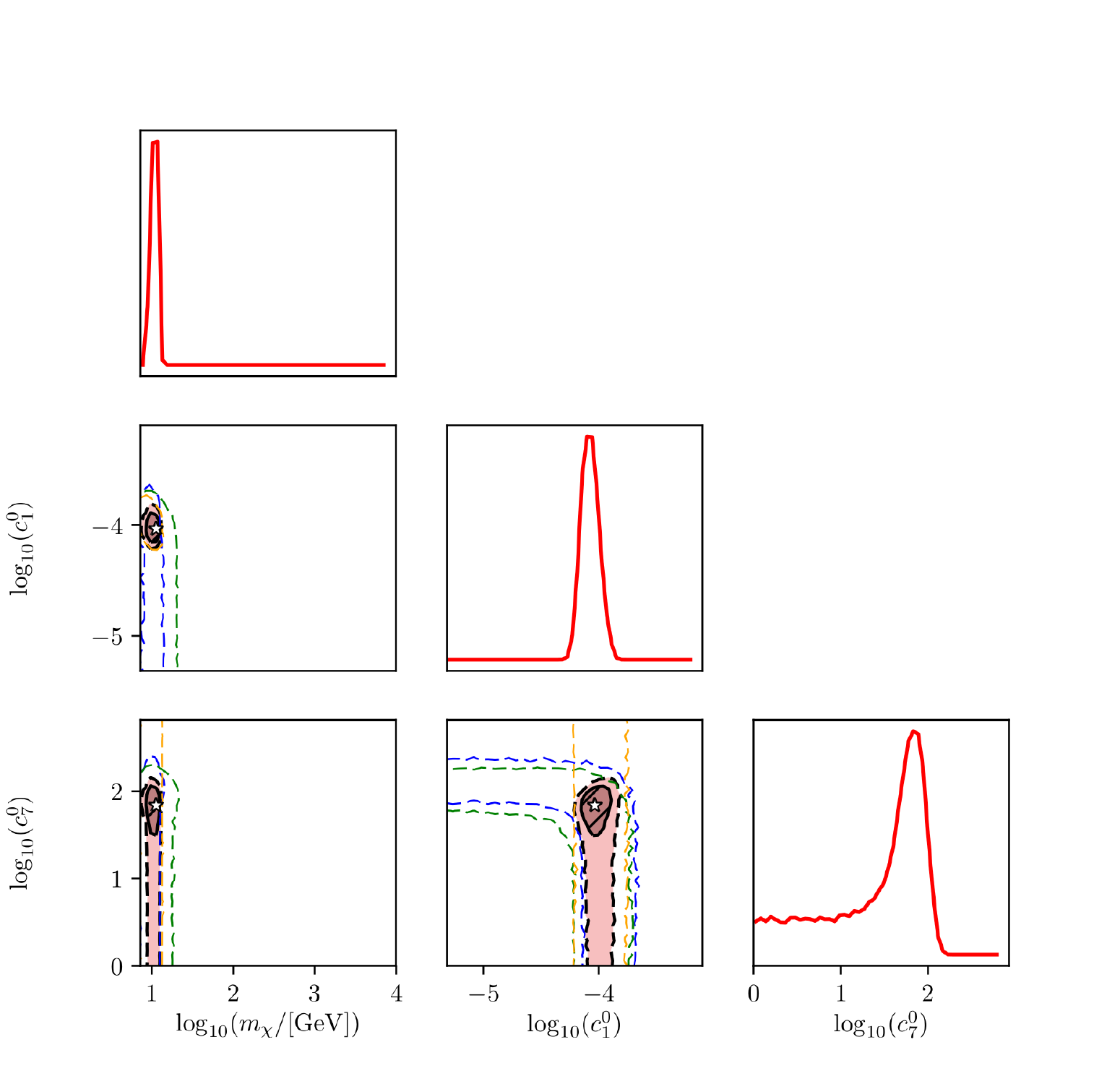}
\end{minipage}
\hspace*{1cm}\begin{minipage}{7cm}
\caption{The same as in Fig.~\ref{fig:bp1ss}, but for simplified model SV.}
\label{fig:bp1sv}
\end{minipage}
\end{figure}

\begin{figure}
\includegraphics[scale=0.55, bb=0 0 892.8 864]{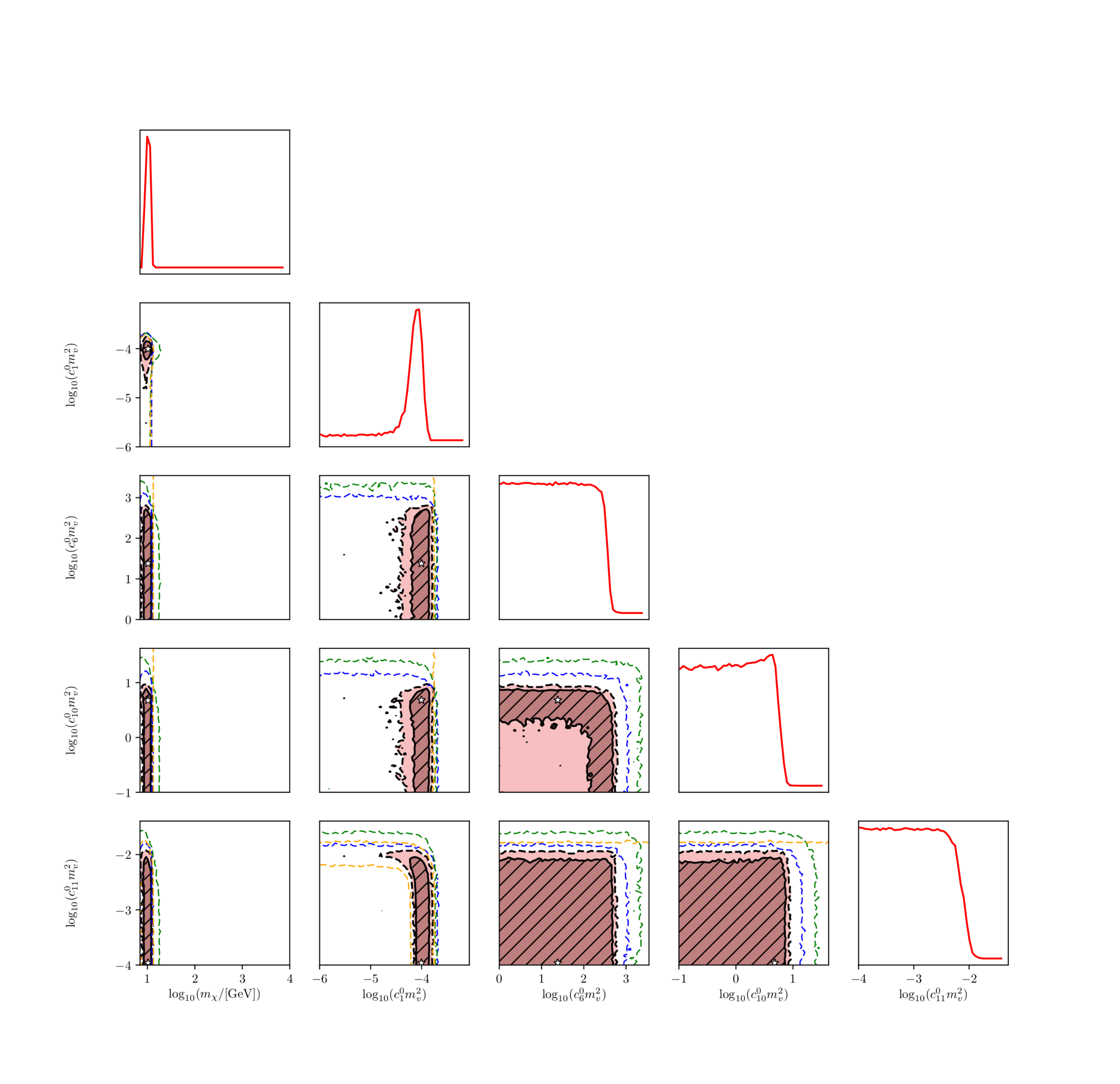}
\caption{The same as in Fig.~\ref{fig:bp1ss}, but for simplified model FS.}
\label{fig:bp1fs}
\end{figure}

\begin{figure}
\includegraphics[scale=0.55, bb=0 0 892.8 864]{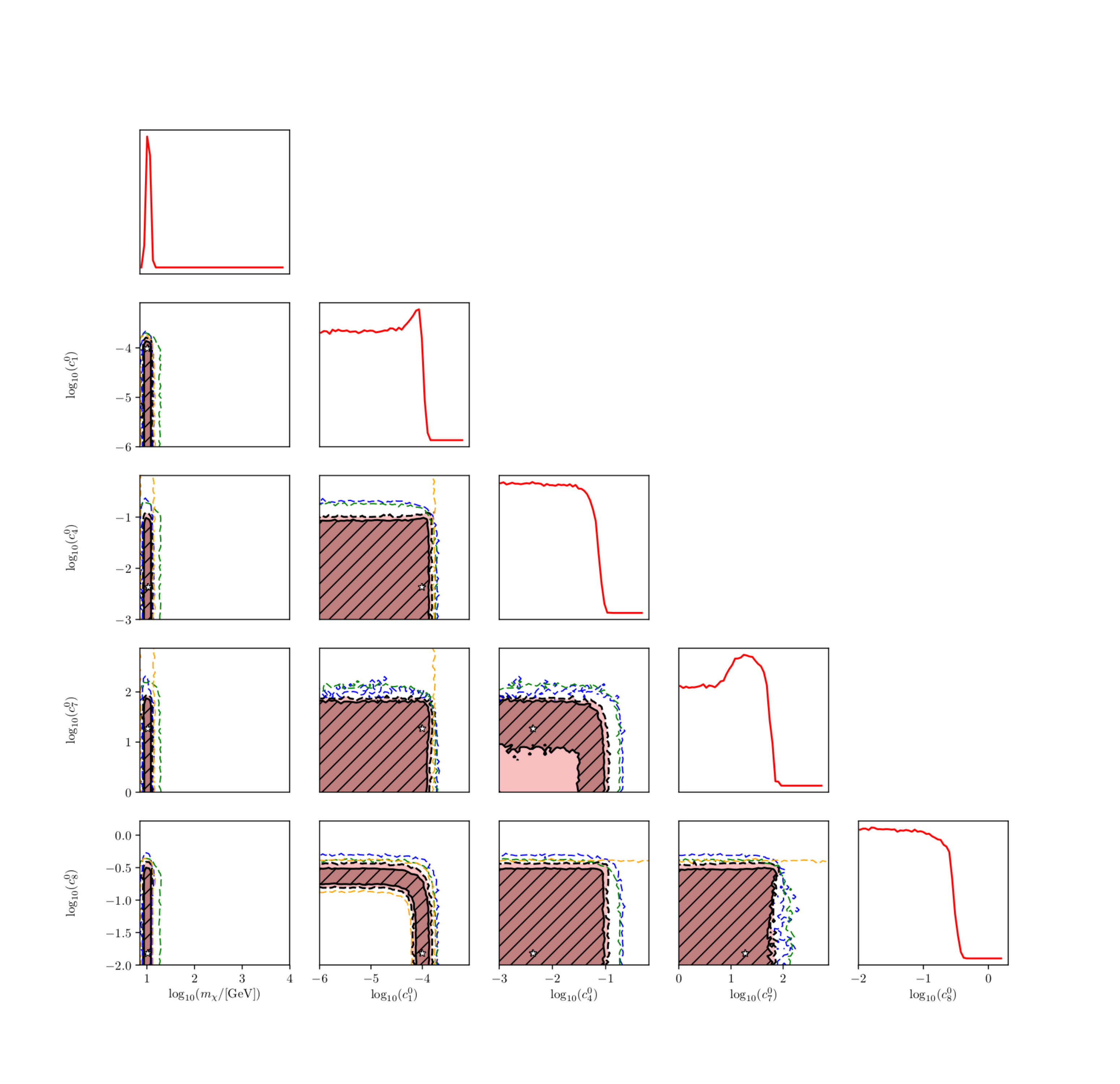}
\caption{The same as in Fig.~\ref{fig:bp1ss}, but for simplified model FV.}
\label{fig:bp1fv}
\end{figure}

\clearpage


\subsection{BP 2}
\label{sec:bp2}

\begin{figure}[h!]
\begin{minipage}{7cm}
\includegraphics[scale=0.52, bb=0 0 446.4 432]{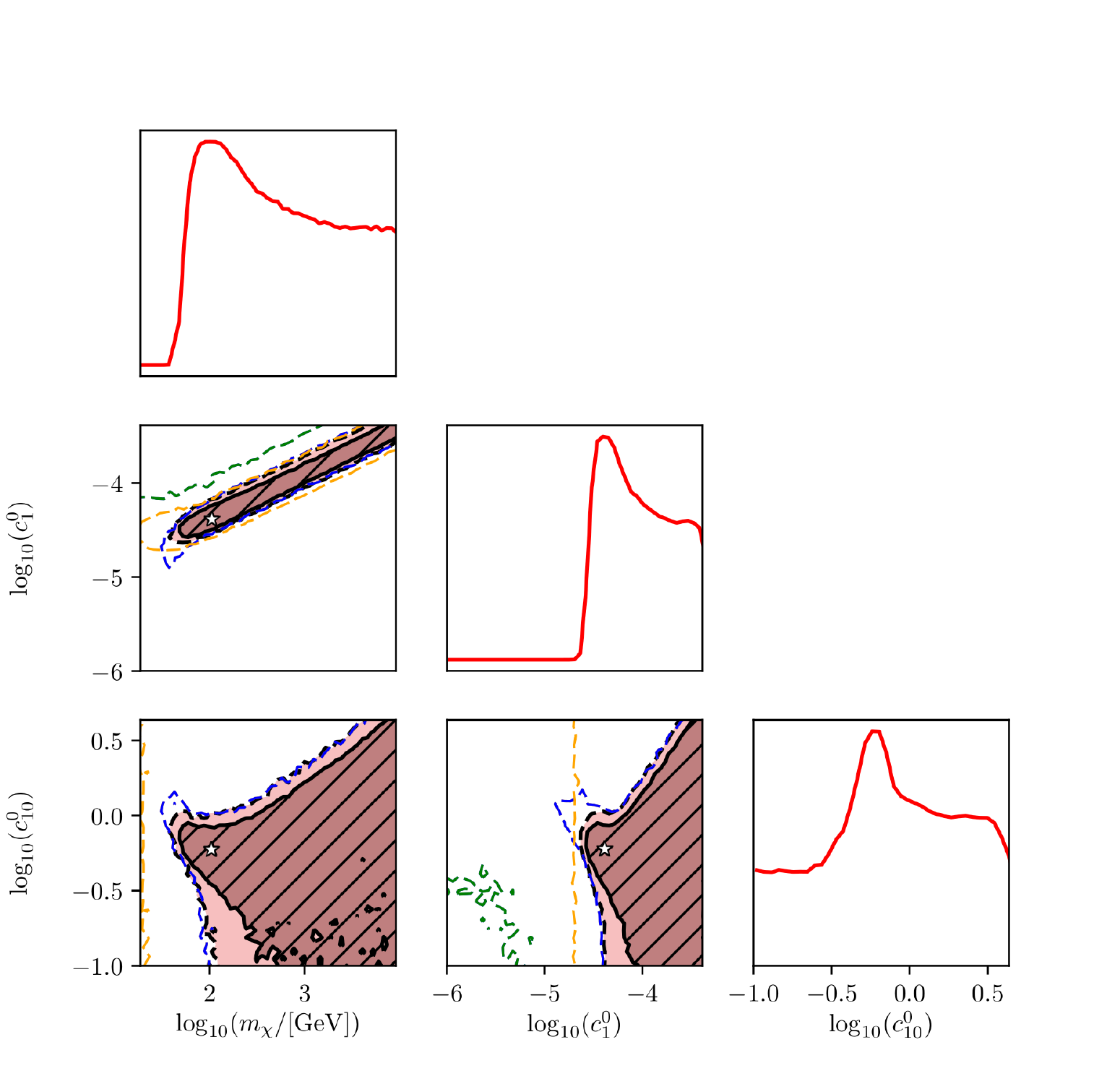}
\end{minipage}
\hspace*{1cm}\begin{minipage}{7cm}
\caption{Profile likelihood for the reconstruction of DM parameters for the simulated data of benchmark point BP2, using simplified model SS. Dashed blue, green, and orange lines correspond to the $2\,\sigma$ (95\% C.L.) contours obtained for individual targets of xenon, germanium, and argon. Black dashed and solid lines correspond to the contours obtained for the combination of the three targets. The best fit point is represented by an asterisk. For reference, the one-dimensional profile likelihoods are also shown.}
\label{fig:bp2ss}
\end{minipage}
\end{figure}

\begin{figure}[h!]
\begin{minipage}{7cm}
\includegraphics[scale=0.52, bb=0 0 446.4 432]{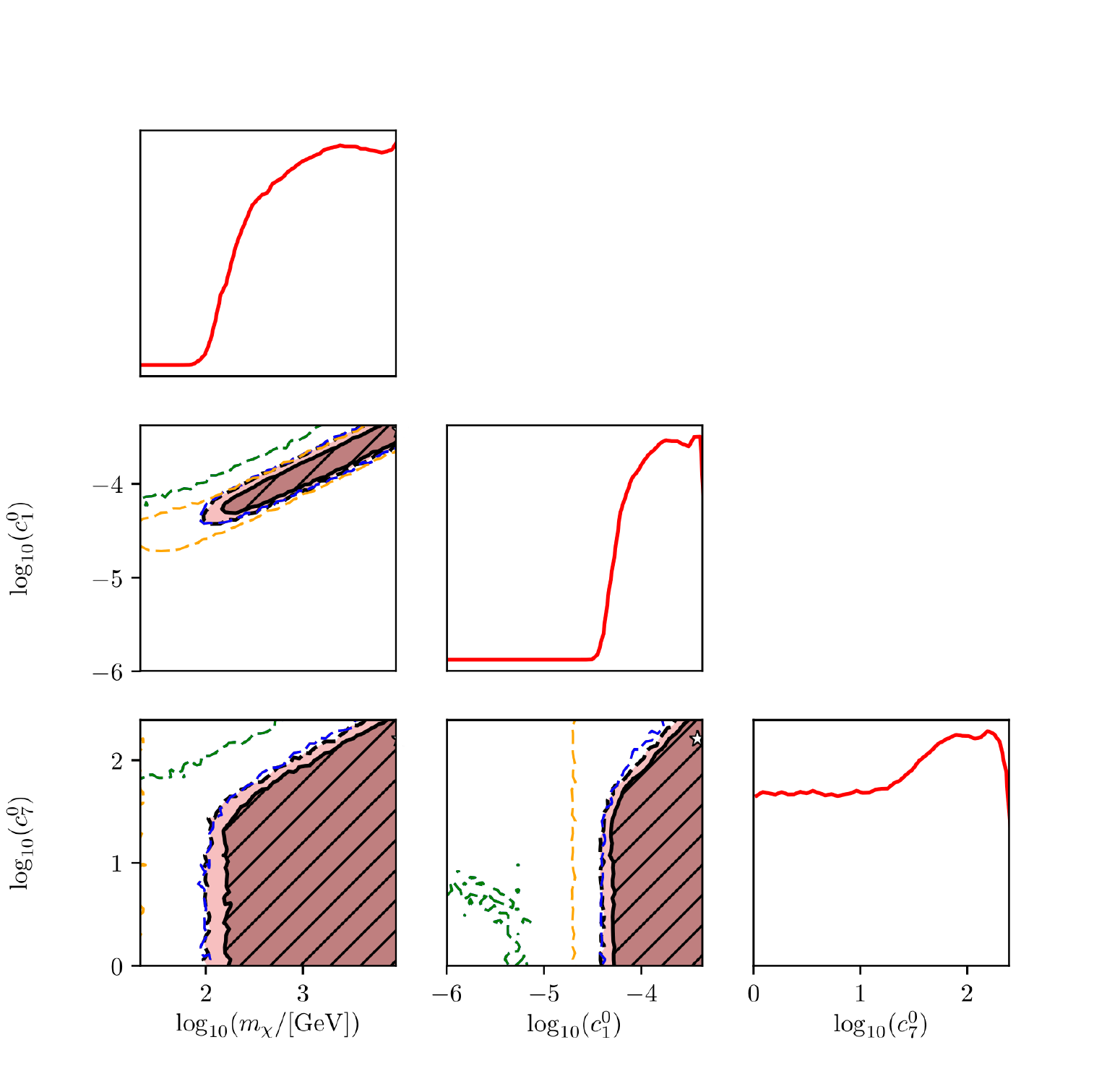}
\end{minipage}
\hspace*{1cm}\begin{minipage}{7cm}
\caption{The same as in Fig.~\ref{fig:bp2ss}, but for simplified model SV.}
\label{fig:bp2sv}
\end{minipage}
\end{figure}

\begin{figure}
\includegraphics[scale=0.55, bb=0 0 892.8 864]{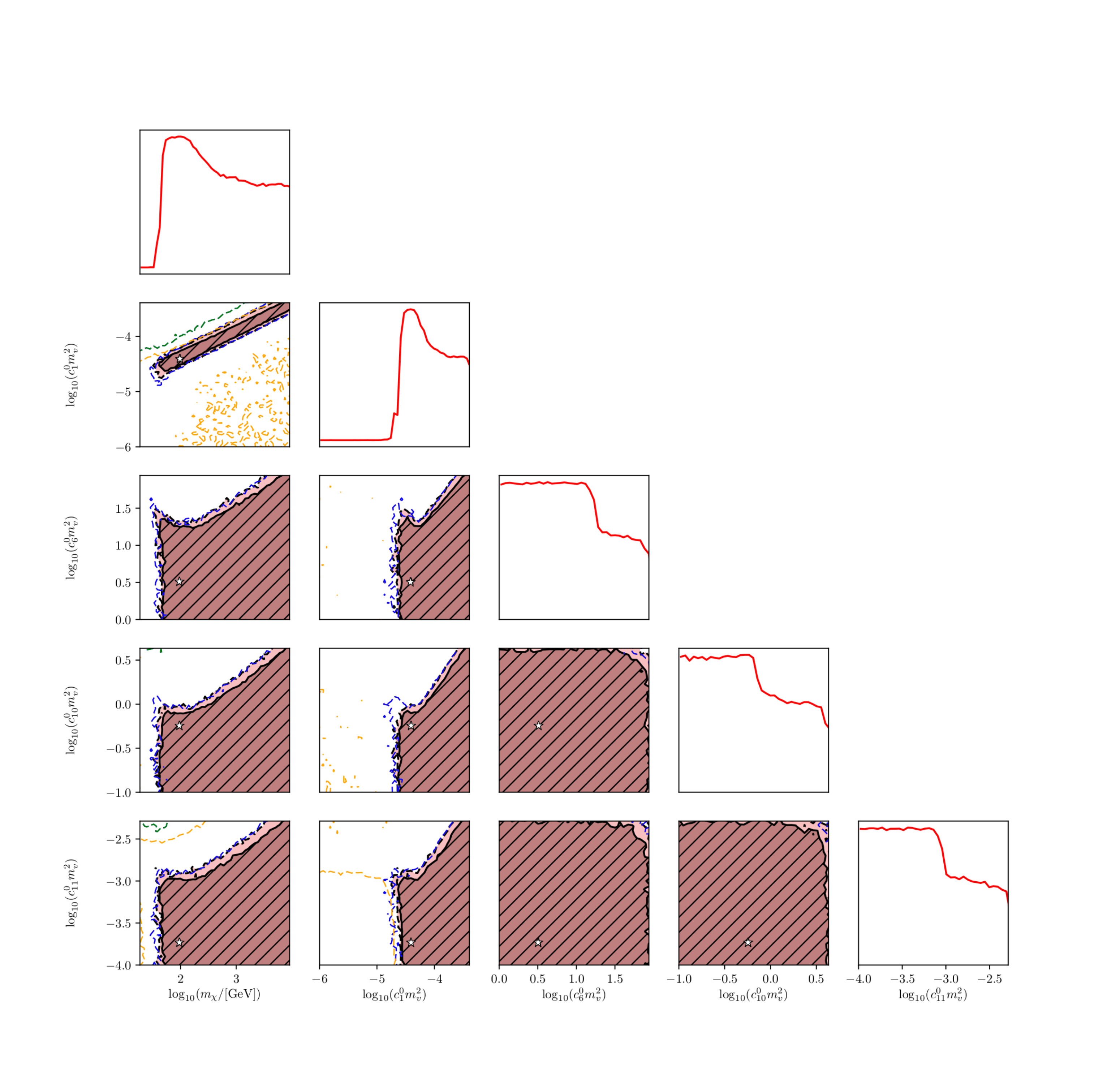}
\caption{The same as in Fig.~\ref{fig:bp2ss}, but for simplified model FS.}
\label{fig:bp2fs}
\end{figure}

\begin{figure}
\includegraphics[scale=0.55, bb=0 0 892.8 864]{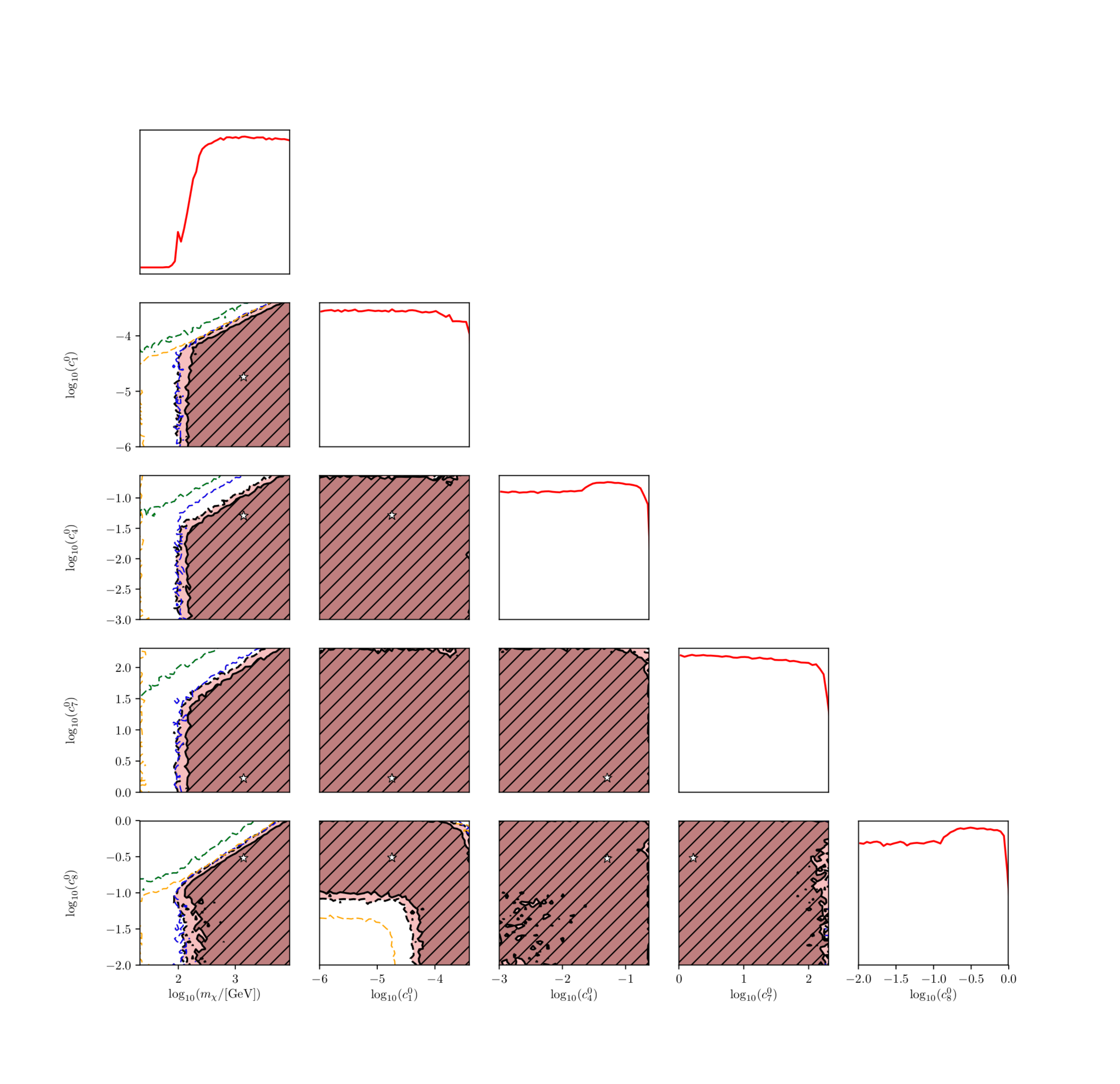}
\caption{The same as in Fig.~\ref{fig:bp2ss}, but for simplified model FV.}
\label{fig:bp2fv}
\end{figure}

\clearpage


\subsection{BP 3}
\label{sec:bp3}

\begin{figure}[h!]
\begin{minipage}{7cm}
\includegraphics[scale=0.52, bb=0 0 446.4 432]{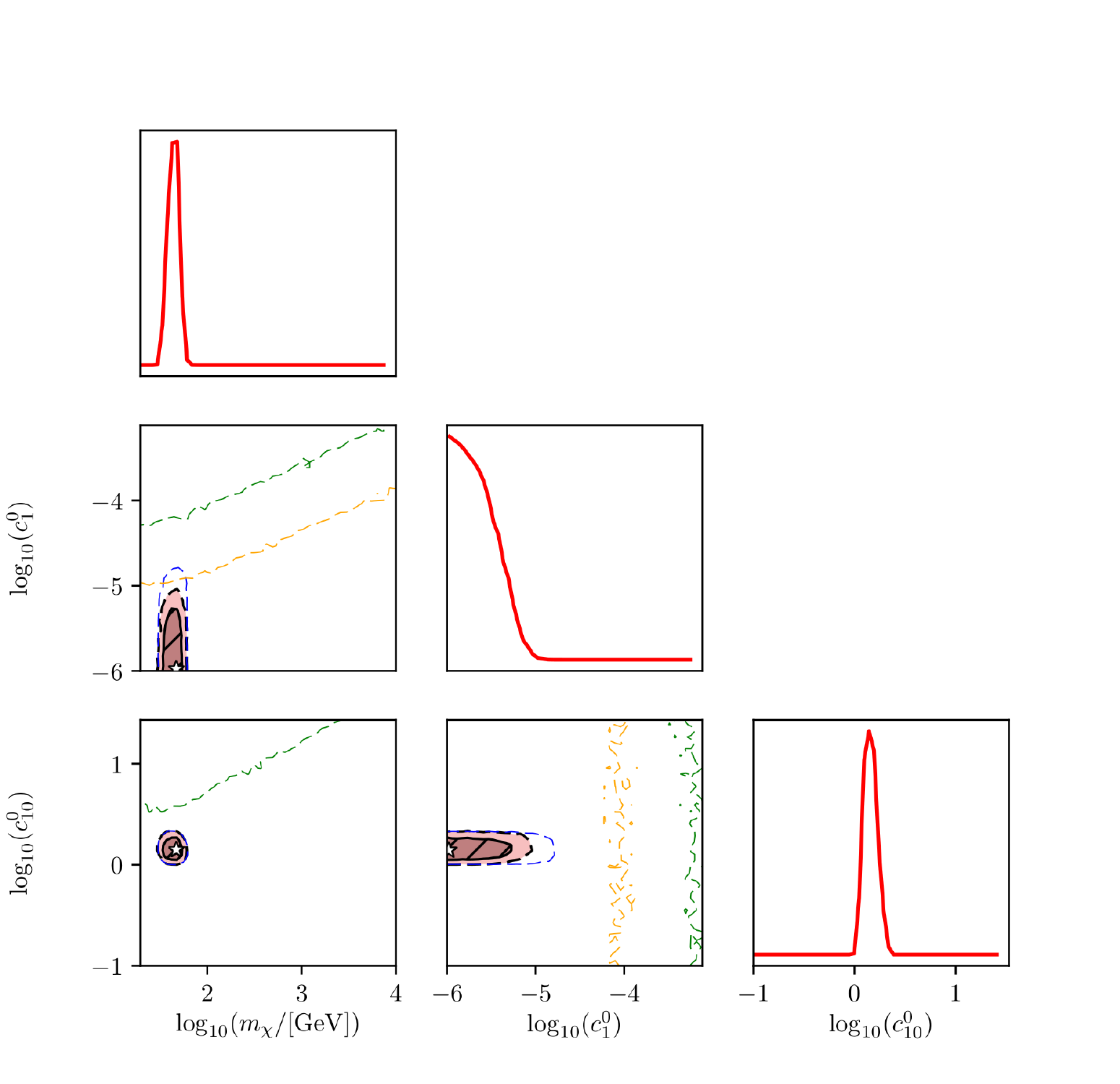}
\end{minipage}
\hspace*{1cm}\begin{minipage}{7cm}
\caption{Profile likelihood for the reconstruction of DM parameters for the simulated data of benchmark point BP3, using simplified model SS. Dashed blue, green, and orange lines correspond to the $2\,\sigma$ (95\% C.L.) contours obtained for individual targets of xenon, germanium, and argon. Black dashed and solid lines correspond to the contours obtained for the combination of the three targets. The best fit point is represented by an asterisk. For reference, the one-dimensional profile likelihoods are also shown.}
\label{fig:bp3ss}
\end{minipage}
\end{figure}

\begin{figure}[h!]
\begin{minipage}{7cm}
\includegraphics[scale=0.52, bb=0 0 446.4 432]{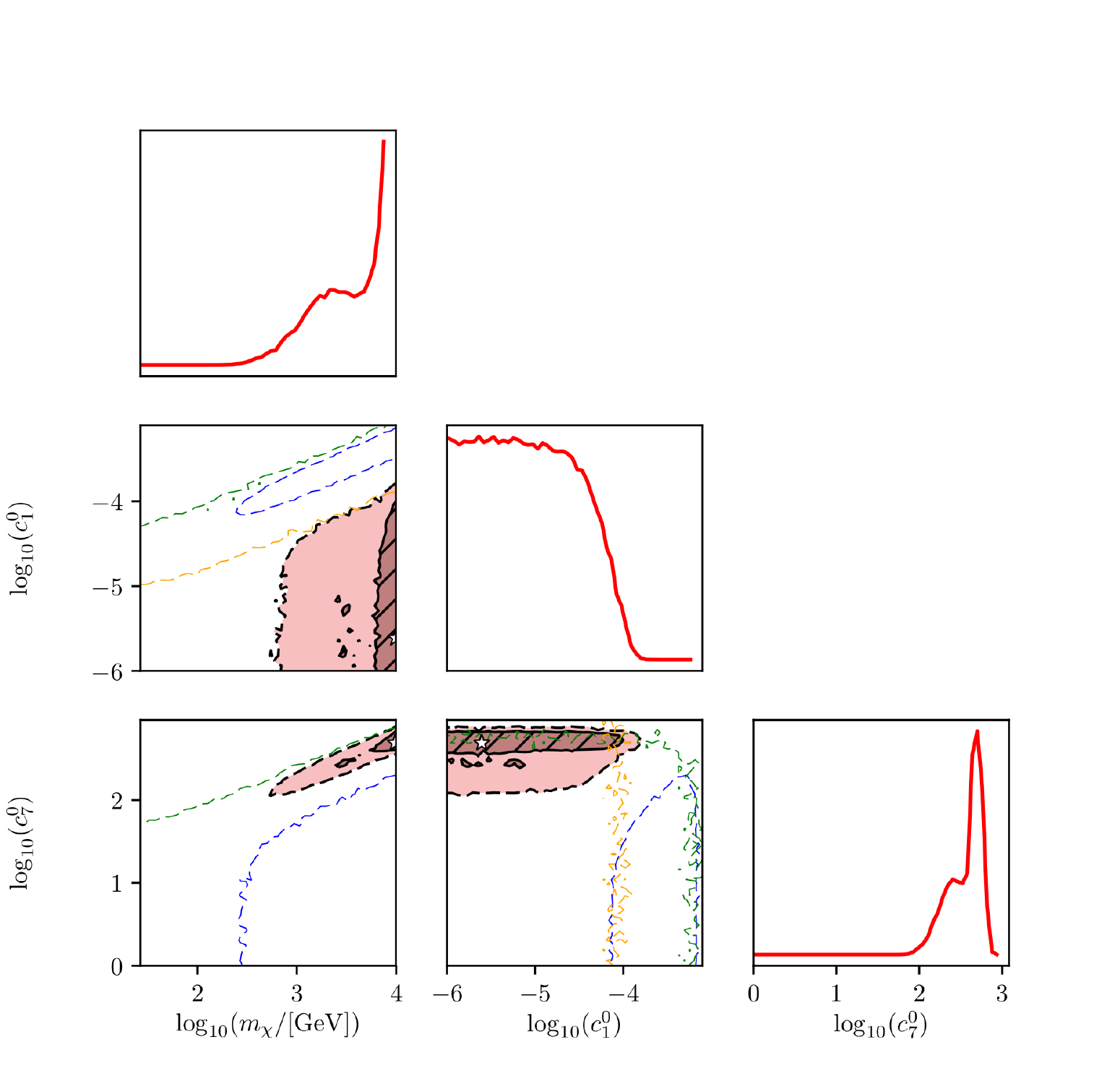}
\end{minipage}
\hspace*{1cm}\begin{minipage}{7cm}
\caption{The same as in Fig.~\ref{fig:bp3ss}, but for simplified model SV.}
\label{fig:bp3sv}
\end{minipage}
\end{figure}

\begin{figure}
\includegraphics[scale=0.55, bb=0 0 892.8 864]{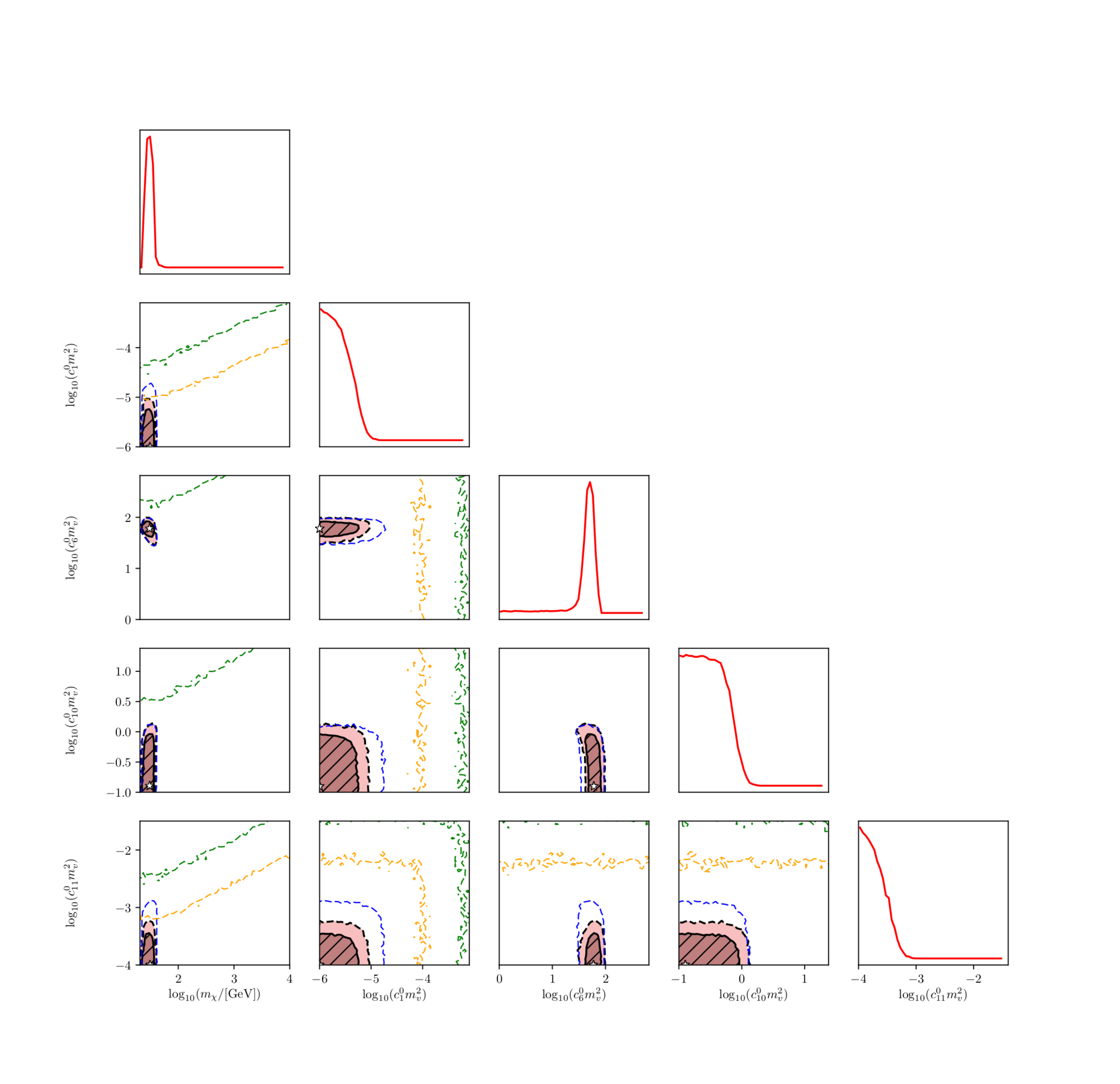}
\caption{The same as in Fig.~\ref{fig:bp3ss}, but for simplified model FS.}
\label{fig:bp3fs}
\end{figure}

\begin{figure}
\includegraphics[scale=0.55, bb=0 0 892.8 864]{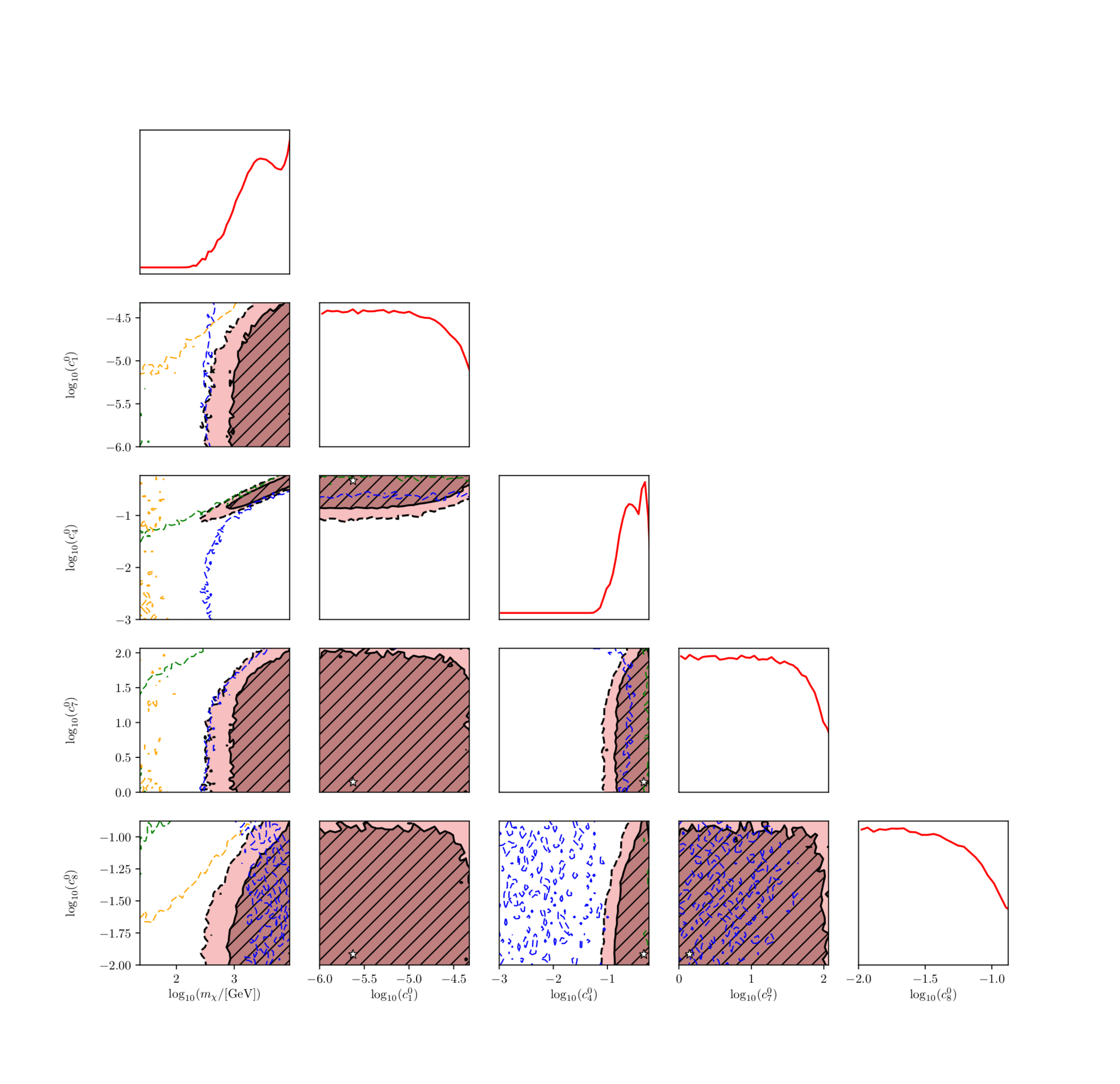}
\caption{The same as in Fig.~\ref{fig:bp3ss}, but for simplified model FV.}
\label{fig:bp3fv}
\end{figure}

\end{document}